\documentclass[twocolumn,showpacs,amsmath,amssymb,superscriptaddress]{revtex4-1}
\usepackage{dcolumn,braket,color,amsmath,footnote,hyperref}
\usepackage{graphicx,bm,url,longtable}
\usepackage{tabularx,multirow,braket}
\newcommand\+{\dagger}

\usepackage{longtable}	% required for `\longtable' (yatex added)
\begin{document}

\title{$\beta$ decay of odd-A nuclei with the interacting boson-fermion model based on the Gogny energy density functional}

\author{K.~Nomura}
\email{knomura@phy.hr}
\affiliation{Department of Physics, Faculty of Science, University of
Zagreb, HR-10000 Zagreb, Croatia}

\author{R.~Rodr\'iguez-Guzm\'an}
\affiliation{Physics Department, Kuwait University, 13060 Kuwait, Kuwait}

\author{L.~M.~Robledo}
\affiliation{Center for Computational Simulation,
Universidad Polit\'nica de Madrid,
Campus de Montegancedo, Boadilla del Monte, 28660-Madrid, Spain
}

\affiliation{Departamento de F\'\i sica Te\'orica, Universidad
	Aut\'onoma de Madrid, E-28049 Madrid, Spain}
\date{\today}

\begin{abstract}
The low-energy excitations and $\beta$ decays   
of  odd-A nuclei are studied within the 
interacting boson-fermion 
model (IBFM), based on the Gogny-D1M nuclear energy density functional (EDF). The
constrained 
Hartree-Fock-Bogoliubov (HFB) approximation is employed to compute 
potential energy surfaces 
in terms of triaxial quadrupole degrees of 
freedom for  even-even Xe and Ba nuclei in the mass 
$A\approx 130$ region. The mean field approximation also provides
spherical single-particle energies and 
occupation probabilities 
for the neighboring odd-A nuclei.
Those quantities represent a microscopic input for 
spectroscopic calculations in 
odd-A Xe and Ba, Cs and La isotopes.
The Gamow-Teller (GT) and Fermi (F) transition matrix elements, 
needed to compute $\beta$-decay  $\log{ft}$ values are 
obtained without any phenomenological fitting. It is shown
that both the 
low-lying states and  $\beta$ 
decays of the studied odd-A systems are described reasonably well
within the employed theoretical framework. 
\end{abstract}


\maketitle

% ----------------------------------------------------------------------

\section{Introduction}

% ----------------------------------------------------------------------

Understanding the structure of the atomic nucleus 
is essential to accurately model fundamental 
processes, such as the $\beta$  and $\beta\beta$ 
decays, and often provides useful insight into  
other domains of physics, 
e.g., it has potential impact on the search for new physics 
beyond the Standard Model of elementary particles. 
Experiments have already been performed at major 
radioactive-ion-beam facilities around the world 
to measure the $\beta$-decay half-lives of numerous 
neutron-rich heavy nuclei  
\cite{dillmann2003,nishimura2011,quinn2012,lorusso2015,caballero2016}. 
 Those  experiments are not only useful for a better understanding 
of nuclear structure phenomena at extreme neutron-to-proton
ratios $N/Z$, but are also instrumental to model the creation of chemical 
elements in various
astrophysical nucleosynthesis scenarios. 
In addition, $\beta$-decay properties are expected 
to be sensitive to  details of the wave functions of low-lying states 
of both the parent and daughter nuclei. Therefore, they also 
serve as a stringent test of various 
nuclear structure models \cite{langanke2003,suhonen2017}.
A number of theoretical calculations have been 
performed to study  $\beta$ decay properties, e.g., 
in terms of the quasiparticle random phase approximation
(QRPA) 
at various levels of sophistication 
\cite{sarriguren2001,simkovic2013,pirinen2015,marketin2016,mustonen2016,nabi2016}, 
the large-scale nuclear shell model 
\cite{langanke2003,caurier2005,honma2005,syoshida2018}, 
and the interacting boson model 
\cite{dellagiacoma1988phdthesis,navratil1988,IBFM,yoshida2002,zuffi2003,brant2004,brant2006,mardones2016}.

The microscopic description of the spectroscopic properties of 
medium-heavy and heavy nuclei is a highly demanding computational
task. In particular, an accurate
description of the excitation spectra 
and transition rates in 
odd-mass and/or odd-odd nuclei
stills remains 
a major challenge in today's nuclear structure theory.
What complicates the microscopic 
description of nuclear systems with  unpaired
nucleons are features such as, for example, the weakening of
pairing correlations, the increase of level densities around the Fermi level, 
polarization effects and  the breaking of time 
reversal symmetry in the intrinsic wave functions 
\cite{RS,bender2003,Rob19,bally2014}.  
Spectroscopic studies of odd-mass and/or odd-odd nuclei has also been carried 
out within the framework of the symmetry-projected Generator Coordinate
Method (GCM) \cite{bally2014,borrajo2016}. However, from a computational 
point of view, this kind of approach is very costly if not 
impossible to apply in heavy nuclei, especially when  
many valence nucleons are involved and/or multiple shape degrees of freedom 
need to be taken into account
in the symmetry-projected GCM ansatz.

An alternative and numerically  feasible EDF-based particle-core coupling
approach to the spectroscopy of odd-A nuclei has been 
developed in previous works \cite{nomura2016odd,nomura2017odd-2}.
Within this approach, the  $(\beta,\gamma)$ potential energy surface 
(PES) of a given even-even core nucleus is computed 
microscopically using the constrained mean field 
approximation.  The mean field approximation also
provides 
the spherical single-particle energies and 
occupation numbers for unpaired nucleon(s) in 
the neighboring odd-A or odd-odd 
nucleus. Those mean field quantities represent 
an essential input to build the Hamiltonian of the 
interacting boson-fermion model (IBFM) 
\cite{iachello1979,IBFM}. 
Three coupling constants for the boson-fermion interaction 
terms are determined so as to reproduce reasonably well 
the experimental low-energy 
spectrum in a given odd-A system. 
At the cost of having to determine these few coupling 
constants empirically, the method allows a detailed 
and simultaneous description of spectroscopy 
in even-even, odd-A, and odd-odd  
nuclei \cite{nomura2019dodd}.

It is then interesting to examine whether the IBFM 
framework based on the microscopic EDF approach can provide at the same time
a consistent description of the low-lying states and $\beta$-decay 
properties in heavy odd-A nuclei. 
In this work, we study those $\beta$ decays where 
only Gamow-Teller (GT) and Fermi (F) transitions are involved, that 
is, the spin of a parent nucleus changes according to 
$\Delta I=0,\pm 1$ and  parity is conserved. 
The $\beta$ decay $\log{ft}$ values require the computation 
of the GT and F transition strengths
which can be obtained using the IBFM 
wave functions for the parent and daughter odd-A nuclei.
One of the advantages of our approach is that the  IBFM
$\beta$ decay calculations do not involve any 
free parameter associated to the GT and F operators. Therefore, $\beta$ decay properties can be  
considered a very stringent test for the IBFM wave functions.

In this work, we focus on the $\beta$ decay of  
odd-A nuclei in the $A\approx 130$ mass region. 
In \cite{nomura2017odd-3}, we have already   
applied the  method to describe 
the $\gamma$-soft-to-near-spherical shape phase 
transitions in  odd-A Xe, Cs, Ba, and La isotopic chains. 
This mass region is one of the most studied in the nuclear chart  
in the context of the $\beta$ decay. Moreover, 
nuclei with $A\approx 130$ exhibit a variety of nuclear structure 
phenomena such as the relevance of triaxial deformations and the existence of 
quantum phase transitions from 
prolate  to $\gamma$-soft (or O(6) limit of IBM \cite{IBM})
and near spherical shapes as one approaches 
the $N=82$ neutron shell closure. The evolution of shapes with neutron number  
might be expected to 
play a role in the corresponding $\beta$ decay 
properties. Phenomenological IBFM studies 
of the low-lying 
states \cite{arias1985} and 
$\beta$ decay properties \cite{zuffi2003,mardones2016}
have already been performed in the same mass
region.

The paper is outlined as follows. In Sec.~\ref{sec:model} we briefly describe the 
procedures followed to 
build the IBFM Hamiltonians from 
the constrained Gogny-HFB calculations. We use the 
parametrization D1M
\cite{D1M} of the Gogny-EDF \cite{Gogny,Rob19} because previous 
studies using the EDF-to-IBM mapping procedure, have shown that
the Gogny-D1M EDF provides a reasonable 
description of the spectroscopic properties 
in medium-mass, heavy odd-A and odd-odd nuclei 
\cite{nomura2017odd-2,nomura2017odd-3,nomura2019dodd,nomura2019cs}
in a wide range of nuclei. In the same section, we also
introduce the $\beta$ decay operators. The results 
of our calculations for the low-lying energy levels in the even-even Xe and Ba nuclei
are discussed in Sec.~\ref{sec:ee}. 
Spectroscopic results from the IBFM calculations for 
the odd-A Xe, Cs, Ba, and La isotopes are presented 
in Sec.~\ref{sec:oe}. The $\log{ft}$ values
obtained for the $\beta$ decays of the studied 
nuclei are discussed in Sec.~\ref{sec:beta}.
Finally, Sec.~\ref{sec:summary} is devoted to the concluding 
remarks.

% ----------------------------------------------------------------------

\section{Theoretical framework\label{sec:model}}

% ----------------------------------------------------------------------

\subsection{Hamiltonian}

Let us introduce the  IBFM Hamiltonian for odd-A systems. 
We use the IBMF-2 version of the IBFM (called IBFM-2) we differentiate between
proton and neutron   
degrees of freedom. In the following, we will
simply denote the IBFM-2 as 
IBFM. 
The IBFM Hamiltonian reads: 
\begin{equation}
\label{eq:ham}
 \hat H_\text{} = \hat H_\text{B} + \hat H_\text{F}^\nu + \hat
 H_\text{F}^\pi + \hat
  H_\text{BF}^\nu + H_\text{BF}^\pi. 
\end{equation}
The first term represents the 
neutron-proton IBM (IBM-2)
Hamiltonian \cite{OAI}  used to describe the even-even core nucleus: 
\begin{equation}
\label{eq:ibm2}
 \hat H_{\text{B}} = \epsilon(\hat n_{d_\nu} + \hat n_{d_\pi})+\kappa\hat
  Q_{\nu}\cdot\hat Q_{\pi}.
\end{equation}
Here $\hat n_{d_\rho}=d^\dagger_\rho\cdot\tilde d_{\rho}$ ($\rho=\nu,\pi$) is the
$d$-boson number operator, and $\hat Q_\rho=d_\rho^\dagger s_\rho +
s_\rho^\dagger\tilde d_\rho^\dagger + \chi_\rho(d^\dagger_\rho\times\tilde
d_\rho)^{(2)}$ is the quadrupole operator. 
The parameters of the Hamiltonian are denoted by 
$\epsilon$, $\kappa$, $\chi_\nu$, and $\chi_\pi$. 
The doubly-magic nucleus $^{132}$Sn is taken as the inert 
core for the boson space. The number of neutron $N_{\nu}$ 
and proton $N_{\pi}$ bosons equals the number of 
neutron-hole and proton-particle pairs, respectively \cite{OAI}.

The second and third term in Eq.~(\ref{eq:ham}) represent the
Hamiltonians for the odd neutron and the odd proton, respectively. Its generic form is 
\begin{equation}
\label{eq:ham-f}
 \hat H_\text{F}^\rho = -\sum_{j_\rho}\epsilon_{j_\rho}\sqrt{2j_\rho+1}
  (a_{j_\rho}^\dagger\times\tilde a_{j_\rho})^{(0)}
\end{equation}
with $\epsilon_{j_\rho}$ being the single-particle energy of the odd
nucleon. Here, $j_\rho$ stands for the angular momentum of the single 
nucleon. On the other hand, $a_{j_\rho}^{(\dagger)}$ 
and $\tilde a_{j_\rho}$ represent fermion creation and annihilation 
operators, with $\tilde a_{jm}=(-1)^{j-m}a_{j-m}$.
For the fermion valence space, we consider the full neutron and proton major
shell $N,Z=50-82$, i.e., $3s_{1/2}$, $2d_{3/2}$, $2d_{5/2}$, $1g_{7/2}$, and $1h_{11/2}$ orbitals.

The fourth and fifth term in Eq.~(\ref{eq:ham}), represent the coupling of the odd
neutron and of the odd proton to the IBM-2 core, respectively: 
\begin{equation}
\label{eq:ham-bf}
 \hat H_\text{BF}^\rho = \Gamma_\rho\hat Q_{\rho'}\cdot\hat q_{\rho} 
+
  \Lambda_\rho\hat V_{\rho'\rho} + A_\rho\hat n_{d_{\rho}}\hat n_{\rho}
\end{equation}
where $\rho'\neq\rho$. 
The first, second, and third terms in the equation above are the
quadrupole dynamical, exchange, and monopole terms, respectively. 
The strength parameters are denoted by  
$\Gamma_\rho$, $\Lambda_\rho$, and $A_{\rho}$. 
As in the previous studies 
\cite{scholten1985,arias1986}, we assume that both the dynamical and exchange terms 
are dominated by the interaction between unlike particles (i.e., between
the odd neutron and proton bosons and between the odd proton and neutron
bosons), and that, for the monopole term, the interaction between
like-particles (i.e., between the odd neutron and neutron bosons and between
the odd proton and proton bosons) plays a dominant role. 
In Eq.~(\ref{eq:ham-bf}) $\hat Q_\rho$ is the same bosonic quadrupole operator as 
in the IBM-2 Hamiltonian in Eq.~(\ref{eq:ibm2}). 
The fermionic quadrupole operator $\hat q_\rho$ reads: 
\begin{equation}
\hat q_\rho=\sum_{j_\rho j'_\rho}\gamma_{j_\rho j'_\rho}(a^\+_{j_\rho}\times\tilde
a_{j'_\rho})^{(2)},
\end{equation} 
where $\gamma_{j_\rho
j'_\rho}=(u_{j_\rho}u_{j'_\rho}-v_{j_\rho}v_{j'_\rho})Q_{j_\rho
j'_\rho}$ and  $Q_{j_\rho j'_\rho}=\langle
l\frac{1}{2}j_{\rho}||Y^{(2)}||l'\frac{1}{2}j'_{\rho}\rangle$ represents
the matrix element of the fermionic 
quadrupole operator in the considered single-particle basis.
The exchange term $\hat V_{\rho'\rho}$ in Eq.~(\ref{eq:ham-bf}) reads: 
\begin{align}
\label{eq:Rayner-new-label}
 \hat V_{\rho'\rho} =& -(s_{\rho'}^\+\tilde d_{\rho'})^{(2)}
\cdot
\Bigg\{
\sum_{j_{\rho}j'_{\rho}j''_{\rho}}
\sqrt{\frac{10}{N_\rho(2j_{\rho}+1)}}\beta_{j_{\rho}j'_{\rho}}\beta_{j''_{\rho}j_{\rho}} \nonumber \\
&:((d_{\rho}^\+\times\tilde a_{j''_\rho})^{(j_\rho)}\times
(a_{j'_\rho}^\+\times\tilde s_\rho)^{(j'_\rho)})^{(2)}:
\Bigg\} + (H.c.), \nonumber \\
\end{align}
with $\beta_{j_{\rho}j'_{\rho}}=(u_{j_{\rho}}v_{j'_{\rho}}+v_{j_{\rho}}u_{j'_{\rho}})Q_{j_{\rho}j'_{\rho}}$.
In the second line of the above equation the notation $:(\cdots):$ indicates normal ordering. 
The definition of the number operator for the odd fermion 
in the monopole interaction has already been introduced in Eq.~(\ref{eq:ham-f}),

\subsection{Procedure to build the IBFM Hamiltonian}

To build the IBFM Hamiltonian, 
we first carry out constrained 
Hartree-Fock-Bogoliubov (HFB) calculations to obtain the potential energy 
surface (PES), as a function of the quadrupole deformation parameters 
$\beta$ and $\gamma$, for a set of even-even Xe and Ba nuclei. 
For each nucleus, the parameters $\epsilon$, $\kappa$, 
$\chi_\nu$, and $\chi_\pi$ of the boson IBM-2 Hamiltonian are fitted to reproduce the HFB PES
when the IBM-2 enery is computed using the  boson coherent state \cite{ginocchio1980} 
(see, Refs.~\cite{nomura2008,nomura2010}, for details).

Next, the single-particle energies $\epsilon_{j_\nu}$ 
($\epsilon_{j_\pi}$) and occupation 
probabilities $v^2_{j_\nu}$ ($v^2_{j_\pi}$) of the unpaired 
neutron and/or proton are computed 
with the help of Gogny-D1M HFB calculations constrained to zero deformation 
\cite{nomura2017odd-2}. These parameters are used in the 
$\hat H_\mathrm{F}^\nu$ ( $\hat H_\mathrm{F}^\pi$) 
and $\hat H_\mathrm{BF}^\nu$ ($\hat H_\mathrm{BF}^\pi$) Hamiltonians, 
respectively. The optimal values of the strength parameters 
for the boson-fermion Hamiltonian 
$\hat H_\mathrm{BF}^\nu$ ($\hat H_\mathrm{BF}^\pi$), i.e., 
$\Gamma_\nu$, $\Lambda_\nu$, and $A_\nu$ 
($\Gamma_\pi$, $\Lambda_\pi$, and $A_\pi$), are determined 
separately for positive and negative parity states, so as to reproduce 
the experimental low-energy 
levels for each of the considered odd-N Xe and Ba 
(odd-Z Cs and La) isotopes. The values of the IBM-2 parameters and the 
IBFM strengths obtained for the 
studied even-even and odd-A nuclei are given in 
Table~\ref{tab:ibm2para} and Table~\ref{tab:ibfm2para}, respectively. 
The IBFM parameters, shown in Table~\ref{tab:ibfm2para}, are exactly the same 
as the ones employed in Ref.~\cite{nomura2019cs}. 
The spherical single-particle energies and occupation 
probabilities for these odd-A nuclei can be found 
in Ref.~\cite{nomura2017odd-3}. 

\begin{table}[htb!]
 \begin{center}
\caption{\label{tab:ibm2para} 
The adopted parameters of the IBM-2 
Hamiltonian $\hat H_\mathrm{B}$ for the even-even-core nuclei 
$^{124-134}$Xe and $^{126-136}$Ba.}
\begin{ruledtabular}
  \begin{tabular}{ccccc}
   & $\epsilon$ (MeV) & $\kappa$ (MeV) & $\chi_\nu$ & $\chi_\pi$ \\
\hline
$^{124}$Xe & 0.45 & $-0.336$ & 0.40 & $-0.50$ \\
$^{126}$Xe & 0.52 & $-0.323$ & 0.25 & $-0.50$ \\
$^{128}$Xe & 0.62 & $-0.315$ & 0.25 & $-0.55$ \\
$^{130}$Xe & 0.82 & $-0.308$ & 0.38 & $-0.50$ \\
$^{132}$Xe & 0.90 & $-0.250$ & 0.20 & $-0.55$ \\
$^{134}$Xe & 0.98 & $-0.190$ & 0.20 & $-0.60$ \\
$^{126}$Ba & 0.28 & $-0.284$ & 0.12 & $-0.49$ \\
$^{128}$Ba & 0.41 & $-0.286$ & 0.12 & $-0.53$ \\
$^{130}$Ba & 0.52 & $-0.297$ & 0.25 & $-0.55$ \\
$^{132}$Ba & 0.65 & $-0.288$ & 0.25 & $-0.45$ \\
$^{134}$Ba & 0.84 & $-0.278$ & 0.40 & $-0.45$ \\
$^{136}$Ba & 1.00 & $-0.278$ & 0.40 & $-0.45$
  \end{tabular}
  \end{ruledtabular}
 \end{center}
\end{table}

\begin{table}[htb!]
\caption{\label{tab:ibfm2para} Strength parameters of the boson-fermion Hamiltonian
 $\hat H_\text{BF}^\rho$ (in MeV) employed for the studied odd-A nuclei. }
 \begin{center}
 \begin{ruledtabular}
  \begin{tabular}{cccc}
      & $\Gamma_\rho$ & $\Lambda_\rho$ & $A_\rho$ \\
\hline
$^{123}$Xe & 3.20 & 0.20 & $-0.14$ \\
$^{125}$Xe & 3.00 & 0.40 & $-0.12$ \\
$^{127}$Xe & 3.00 & 0.60 & $-0.28$ \\
$^{129}$Xe & 1.60 & 2.20 & $-0.30$ \\
$^{131}$Xe & 1.00 & 2.00 & $-0.30$ \\
$^{133}$Xe & 0.30 & 2.00 & $-0.30$ \\
$^{125}$Ba & 3.00 & 1.55 & $0.0$ \\
$^{127}$Ba & 3.00 & 0.60 &  $-0.35$ \\
$^{129}$Ba & 1.60 & 1.50 & $-0.80$ \\
$^{131}$Ba & 1.20 & 1.80 & $-0.50$ \\
$^{133}$Ba & 1.00 & 1.60 & $-0.55$ \\
$^{135}$Ba & 0.30 & 1.60 & $-0.50$ \\
$^{125}$Cs & 0.80 & 0.51 & $-0.80$ \\
$^{127}$Cs & 0.80 & 0.40 & $-0.70$ \\
$^{129}$Cs & 1.00 & 0.40 & $-0.70$ \\
$^{131}$Cs & 1.20 & 0.55 & $-0.80$ \\
$^{133}$Cs & 1.20 & 0.58 & $-0.50$ \\
$^{135}$Cs & 0.80 & 1.00 & $-0.10$ \\
$^{127}$La & 1.00 & 1.50 & $-2.7$ \\
$^{129}$La & 0.80 & 1.76 & $-2.0$ \\
$^{131}$La & 0.80 & 1.92 & $-2.3$ \\
$^{133}$La & 1.00 & 2.00 & $-1.1$ \\
$^{135}$La & 1.50 & 0.81 & $-0.45$ \\
$^{137}$La & 2.00 & 1.45 & $0.0$ 
  \end{tabular}
  \end{ruledtabular}
 \end{center}
\end{table}

The resulting IBFM Hamiltonian is then diagonalized 
in the basis $\ket{[L_{\nu}\otimes L_{\pi}]^{(L)}\otimes j_{\rho}]^{(I)}} $, 
where $L_\rho$ is the angular momentum 
of the neutron or proton boson system, $L$ is the total angular 
momentum of the boson system, 
and $I$ represents the total angular momentum of the coupled 
boson-fermion system.

\subsection{Electromagnetic transition operators}

The electromagnetic transition rates in odd-A nuclei 
can be computed using the eigenstates of the IBFM Hamiltonian. 
Here, we consider the electric quadrupole (E2) and magnetic dipole (M1) 
properties. The E2 operator $\hat T^\mathrm{(E2)}$ reads 
\cite{zuffi2003,nomura2019cs}:  
\begin{align}
 \label{eq:e2}
\hat T^\mathrm{(E2)}&
= e_\nu^\mathrm{B}\hat Q_\nu + e_\pi^\mathrm{B}\hat Q_\pi
-\frac{1}{\sqrt{5}}\sum_{\rho=\nu,\pi}\sum_{j_{\rho}j'_{\rho}} \nonumber \\
&\times(u_{j_{\rho}}u_{j'_{\rho}}-v_{j_{\rho}}v_{j'_{\rho}})\langle
j'_{\rho}||e^\mathrm{F}_{\rho}r^2Y^{(2)}||j_{\rho}\rangle(a_{j_{\rho}}^\dagger\times\tilde a_{j'_{\rho}})^{(2)},
\nonumber \\
\end{align}
where $e^\mathrm{B}_\rho$ and $e^\mathrm{F}_{\rho}$
are the effective charges for the boson and fermion systems,
respectively. We have used the fixed values 
$e^\mathrm{B}_\nu=e^\mathrm{B}_\pi=0.108$ $e$b, and $e^\mathrm{F}_\nu=0.5$ $e$b and $e^\mathrm{F}_\pi=1.5$ $e$b. These values have already been employed in previous 
IBFM calculations \cite{zuffi2003} 
for the same mass region. 
The M1 transition operator $\hat T^\mathrm{(M1)}$ reads 
\cite{zuffi2003,nomura2019cs}: 
\begin{align}
 \label{eq:m1}
\hat T^\mathrm{(M1)}&=\sqrt{\frac{3}{4\pi}}
\Big\{
g_\nu^\mathrm{B}\hat L^\mathrm{B}_\nu + g_\pi^\mathrm{B}\hat
L^\mathrm{F}_\pi
-\frac{1}{\sqrt{3}}\sum_{\rho=\nu,\pi}\sum_{j_\rho j_\rho'} \nonumber \\
&\times (u_{j_{\rho}}u_{j'_{\rho}}+v_{j_{\rho}}v_{j'_{\rho}})\langle
j'_{\rho}||g_l^\rho{\bf l}+g_s^\rho{\bf s}||j_{\rho}\rangle(a_{j_{\rho}}^\dagger\times\tilde
a_{j'_{\rho}})^{(1)}
\Big\}. \nonumber \\
\end{align}
In this expression, $g_\nu^\mathrm{B}$ and 
$g_\pi^\mathrm{B}$ are the $g$-factors for the neutron and
proton bosons, respectively. 
The empirical values $g_\nu^\mathrm{B}=0\,\mu_N$ and
$g_\pi^\mathrm{B}=0.8\,\mu_N$, taken from \cite{zuffi2003}, 
are used for all the studied odd-A nuclei. 
For the neutron (proton) $g$-factors, the usual Schmidt values 
$g_l^\nu=0\,\mu_N$ and $g_s^\nu=-3.82\,\mu_N$
($g_l^\pi=1.0\,\mu_N$ and $g_s^\pi=5.58\,\mu_N$) are used. 
The $g_s$ values, for both protons and neutrons, have been 
quenched by 30\%.

\subsection{Gamow-Teller and Fermi transition operators}

To obtain the $\beta$-decay $\log{ft}$ values, 
the Gamow-Teller (GT) and Fermi (F) 
matrix elements using the wave functions corresponding to the  
initial state (with spin $\ket{I_\mathrm{i}}$) for the parent nucleus 
and the final state (with spin $\ket{I_\mathrm{f}}$) for the daughter nucleus are needed. 
Those wave functions are obtained with two independent IBFM calculations. 
The GT and F operators have to be defined in the boson-fermion space of the IBFM. To this
end we introduce the  one-fermion transfer operators 
\cite{dellagiacoma1988phdthesis}: 
%\begin{widetext}
\begin{align}
\label{eq:creation1}
A^{(j)\dagger}_{m} &= \zeta_{j} a_{jm}^{\dagger}
 + \sum_{j'} \zeta_{jj'} s^{\dagger}_\rho (\tilde{d}_\rho\times a_{j'}^{\dagger})^{(j)}_{m}
\nonumber \\
    & \hspace{3em} (\Delta n_{j} = 1, \; \Delta N_\rho=0)
\end{align}
and
\begin{align}
\label{eq:creation2}
B^{(j)\dagger}_{m} &= \theta_{j} s^{\dagger}_\rho\tilde{a}_{jm}
 + \sum_{j'} \theta_{jj'} (d^{\dagger}_\rho\times\tilde{a}_{j'})^{(j)}_{m}
 \nonumber \\
  &\hspace{3em} (\Delta n_{j} = -1, \; \Delta N_\rho = 1) .
\end{align}
Both operators increase the number of valence 
neutrons (protons) $n_{j} + 2N_\rho$ by one. 
Note, that the index of $j_\rho$ is omitted 
for the sake of simplicity. 
The conjugate operators read:
%\begin{widetext}
\begin{align}
\label{eq:annihilation1}
\tilde{A}^{(j)}_{m}
&= (-1)^{j-m} \left\{ A^{(j)\dagger}_{-m} \right\}^{\dagger} \nonumber \\
&= \zeta_{j}^{*} \tilde{a}_{jm}
+ \sum_{j'} \zeta_{jj'}^{*} s_\rho (d^{\dagger}_\rho\times\tilde{a}_{j'})^{(j)}_{m}
\nonumber  \\
&\hspace{3em} (\Delta n_{j} = -1, \; \Delta N_\rho = 0) 
\end{align}
and
\begin{align}
\label{eq:annihilation2}
\tilde{B}^{(j)}_{m}
&= (-1)^{j-m} \left\{ B^{(j)\dagger}_{-m} \right\}^{\dagger}
\nonumber \\
&= -\theta_{j}^{*} s_\rho a_{jm}^{\dagger}
- \sum_{j'} \theta_{jj'}^{*} (\tilde{d}_\rho\times a_{j'}^{\dagger})^{(j)}_{m}
   \nonumber \\
   &\hspace{1em} (\Delta n_{j} = 1, \; \Delta N_\rho = -1) .
\end{align}
These operators decrease the 
number of valence 
neutrons (protons) $n_{j} + 2N_\rho$ by one.

The coefficients $\zeta_{j}$, $\zeta_{jj'}$, $\theta_{j}$, 
and $\theta_{jj'}$ in Eqs.~(\ref{eq:creation1})-(\ref{eq:annihilation2}) are 
given \cite{IBFM} by
\begin{align}
\label{eq:zeta1}
\zeta_{j} &= u_{j} \frac{1}{K_{j}'} , \\
\label{eq:zeta2}
\zeta_{jj'} &= -v_{j} \beta_{j'j} 
\sqrt{\frac{10}{N_\rho(2j+1)}}\frac{1}{KK_{j}'} , \\
\label{eq:theta1}
\theta_{j} &= \frac{v_{j}}{\sqrt{N_\rho}} \frac{1}{K_{j}''} , \\
\label{eq:theta2}
\theta_{jj'} &= u_{j} \beta_{j'j} \sqrt{\frac{10}{2j+1}} \frac{1}{KK_{j}''} .
\end{align}
The parameters $K$, $K_j'$, and $K_j''$ read 
\cite{dellagiacoma1988phdthesis,IBFM}: 
\begin{subequations}
\begin{align}
&K = \left( \sum_{jj'} \beta_{jj'}^{2} \right)^{1/2},\\
&K_{j}' = \left( 1 + 2 \left(\frac{v_{j}}{u_{j}}\right)^{2}
\frac{\braket{(\hat n_{s_\rho}+1)\hat n_{d_\rho}}_{0^+_1}}
{N_\rho(2j+1)} \frac{\sum_{j'} \beta_{j'j}^{2}}{K^{2}} \right)^{1/2} ,\\
&K_{j}'' = \left(
\frac{\braket{\hat n_{s_\rho}}_{0^+_1}}{N_\rho}
+2\left(\frac{u_{j}}{v_{j}}\right)^{2}
\frac{\braket{\hat n_{d_\rho}}_{0^+_1}}{2j+1}
\frac{\sum_{j'} \beta_{j'j}^{2}}{K^{2}} \right)^{1/2}
\end{align}
\end{subequations}
Note that $\hat n_{s_{\rho}}$ is the number operator 
for the $s_\rho$ boson 
and that $\braket{\cdots}_{0^+_1}$ represents the 
expectation value of a given operator in the 
$0^+_1$ ground state of the considered even-even nucleus. 
For a more detailed account, the reader is referred
to Refs.~\cite{dellagiacoma1988phdthesis,IBFM}.

With the previously defined building blocks, the IBFM images 
of the Fermi ( $\sum_{k} t^{\pm}_{k}$), and 
Gamow-Teller ($\sum_{k} t^{\pm}_{k} {\bf \sigma}_{k}$) transition operators, 
take the form
\begin{align}
\hat{\cal O}^{\rm F} &=-\sum_{j}\sqrt{2j+1} 
\left(P^{(j)}_{\nu}\times P^{(j)}_{\pi}\right)^{(0)}, \\
\hat{\cal O}^{\rm GT} &= \sum_{j'j}
 \eta_{j'j} \left(P^{(j')}_{\nu}\times P^{(j)}_{\pi}\right)^{(1)}
\end{align}
where
\begin{align}
\label{eq:eta}
\eta_{j'j} &= - \frac{1}{\sqrt{3}} \langle\ell' \frac{1}{2} ; j' ||
 {\bf \sigma} || \ell \frac{1}{2} ; j\rangle
 \nonumber \\
&= - \delta_{\ell'\ell} \sqrt{2(2j'+1)(2j+1)}
W\left(\ell j' \frac{1}{2} 1 ; \frac{1}{2} j\right) ,
\end{align}
with $W$ being a Racah coefficient. In the case of  
$\beta^+$ decay, $P_{\nu}^{(j')} = \tilde B_\nu^{(j')}$ 
and $P_{\pi}^{(j)} = \tilde A_\pi^{(j)}$ while 
for $\beta^-$ decay
$P_{\nu}^{(j')} = B_\nu^{(j')\dagger}$ 
and $P_{\pi}^{(j)} = {A_\pi^{(j)\dagger}}$. Then, the
reduced Fermi
$B(\mathrm{F}; I_\mathrm{i}\rightarrow I_\mathrm{f})$
and GT
$B(\mathrm{GT}; I_\mathrm{i}\rightarrow I_\mathrm{f})$
transition rates 
read: 
\begin{align}
B(\mathrm{F}; I_\mathrm{i}\rightarrow I_\mathrm{f})
&=
\frac{1}{2I_\mathrm{i}+1} |\langle I_\mathrm{f} ||\hat{\cal O}^\mathrm{F} || I_\mathrm{i} \rangle |^{2}\\
B(\mathrm{GT}; I_\mathrm{i}\rightarrow I_\mathrm{f})
&=
\frac{1}{2I_\mathrm{i}+1} |\langle I_\mathrm{f} ||\hat{\cal O}^\mathrm{GT} || I_\mathrm{i} \rangle |^{2}
\end{align}
The $\log ft$ for the 
$\beta$ decay $I_\mathrm{i}\rightarrow  I_\mathrm{f}$, 
can be computed as:
\begin{equation}
\label{eq:ft}
\log{ft} = 
\log_{10}\left\{\frac{6163}{B(\mathrm{F};I_\mathrm{i}\rightarrow  I_\mathrm{f})+g_A^2B(\mathrm{GT};I_\mathrm{i}\rightarrow  I_\mathrm{f})}\right\}.
\end{equation}
Here, $g_A$ is the ratio of the axial-vector to vector coupling constants, 
$g_A=G_{A}/G_{V}$. We have employed the free nucleon value  
$g_A=1.2701(25)$ \cite{beringer2012} for all the studied nuclei  
without  quenching.

\section{Even-even nuclei\label{sec:ee}}

% ----------------------------------------------------------------------
%                  Even-even Xe and Ba
% ----------------------------------------------------------------------

\begin{figure}[htb!]
\begin{center}
\includegraphics[width=\linewidth]{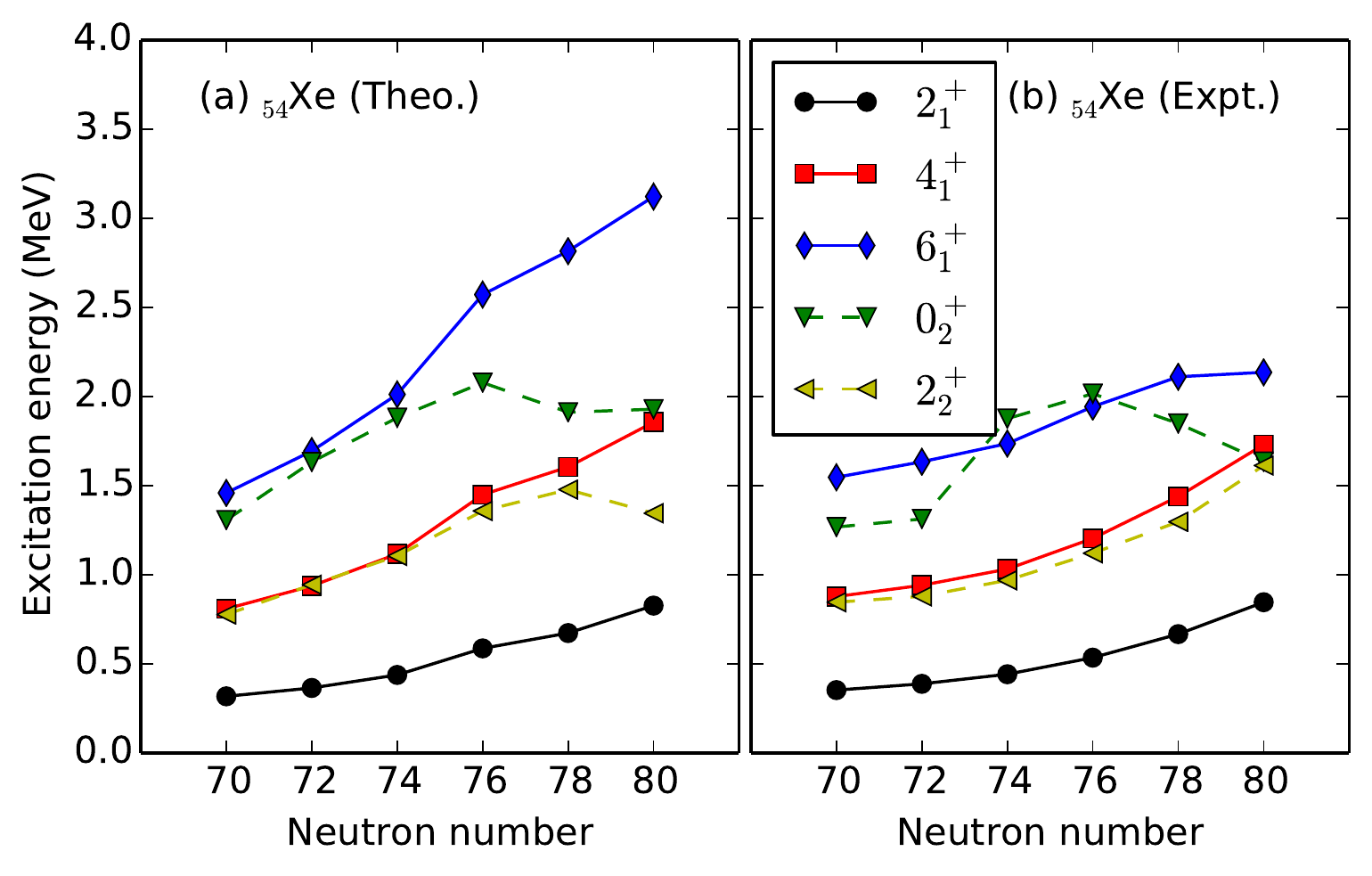}\\
\includegraphics[width=\linewidth]{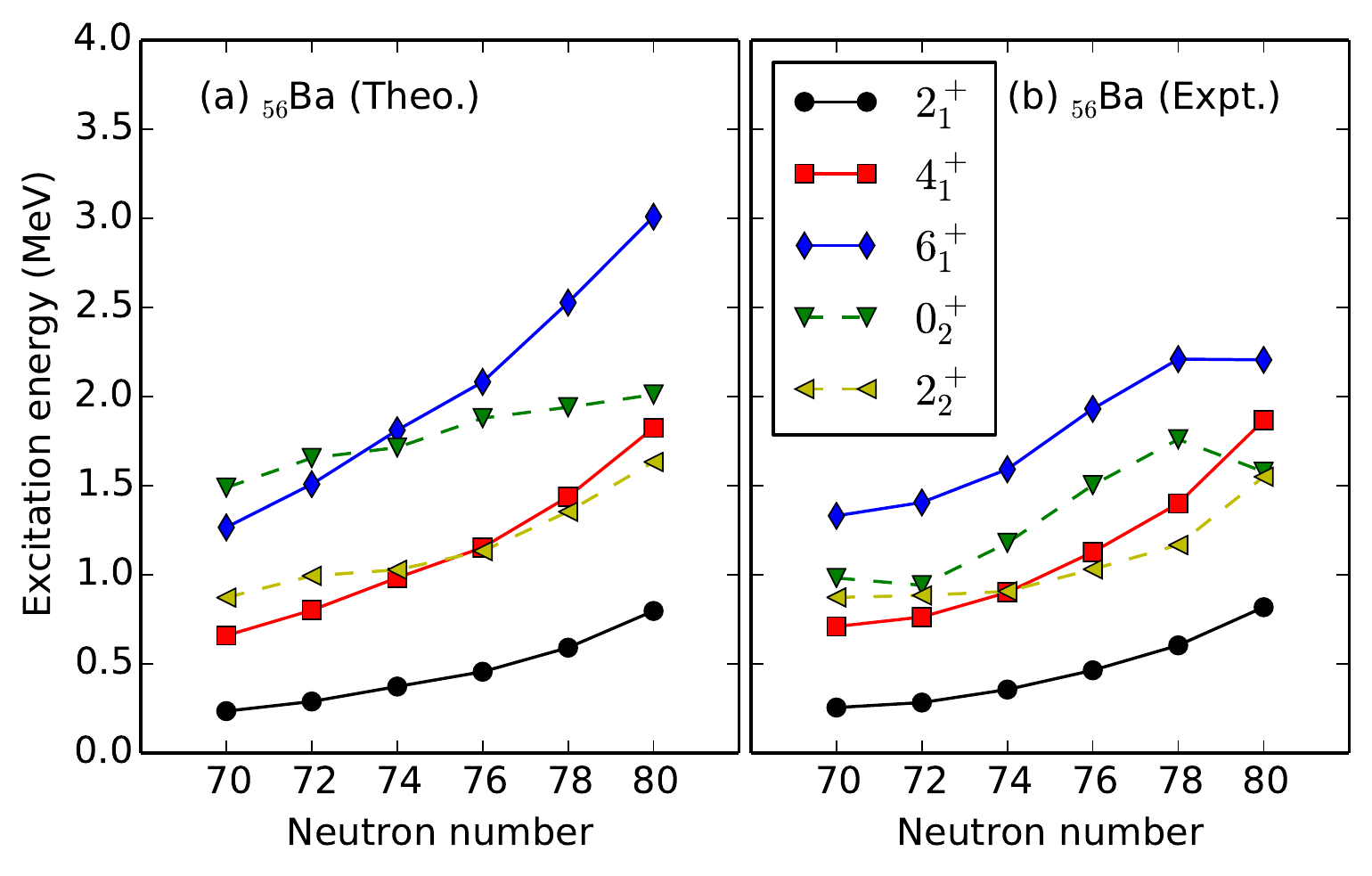}
\caption{(Color online) 
The low-lying excitation spectra, obtained
for the even-even nuclei $^{124-134}$Xe and $^{126-136}$Ba
with the mapped IBM-2 Hamiltonian, are compared with the 
corresponding experimental data \cite{data} taken 
from the  ENSDF database.}
\label{fig:level-even}
\end{center}
\end{figure}

The low-lying excitation spectra, obtained for 
Xe and Ba nuclei with the mapped IBM-2 
Hamiltonian, are compared with the corresponding experimental data \cite{data} 
in Fig.~\ref{fig:level-even}. As can be seen from 
the figure, the predicted spectra reproduce 
well the experimental ones for both the Xe and Ba isotopic 
chain. Note, that the IBM-2 description of the low-lying energy
levels in Xe and Ba nuclei obtained in this work is 
more accurate than in Ref.~\cite{nomura2017odd-3}
where no distinction was made between neutron 
and proton degrees of freedom within the IBM-1 model. 
From  $N=70$ to $N\approx 76$, both the predicted and empirical energy levels exhibit 
features of $\gamma$-soft nuclei (such as the 
the $R_{4/2}$ between the excitation energies of the 
 $4^+_1$ and  $2^+_1$ states close to 2.5, a low-lying 
$2^+_2$ level close 
in energy to the $4^+_1$ level and a 
$0^+_2$ level 
close in energy to the $6^+_1$ one).
The transition from $\gamma$-soft to vibrational 
spectra is characterized by the 
behavior of the $0^+_2$ energy level. It starts to decouple 
from the $6^+_1$ level from 
$N=76$ to $N=78$ (in Xe) or from $N=78$ to 
$N=80$ (in Ba). Triplets are then formed with 
the $4^+_1$ and $2^+_2$ levels, which is a typical 
feature of a multi-phonon spectrum. 

The behavior of 
the excitation energies in  even-even Xe and Ba 
nuclei, as functions of the neutron number $N$,  is consistent 
with the gradual changes observed in the topology of the 
underlying PESs, presented in Fig.~5 of Ref.~\cite{nomura2017odd-3}. 
Those PESs exhibit a prolate or $\gamma$-soft minimum 
for the nuclei with $N\leqslant 76$ (in Xe) and 78 (in Ba),
as well as transitions to nearly spherical shapes 
for $N\geqslant 78$ (in Xe) and $N\geqslant 80$ (in Ba).
For a more detailed account, the reader is referred 
to Ref.~\cite{nomura2017odd-3}.

% ----------------------------------------------------------------------
%                  odd-A Xe, Ba, and Cs
% ----------------------------------------------------------------------

\section{Odd-A nuclei\label{sec:oe}}

\begin{figure*}[htb!]
\begin{center}
\includegraphics[width=0.7\linewidth]{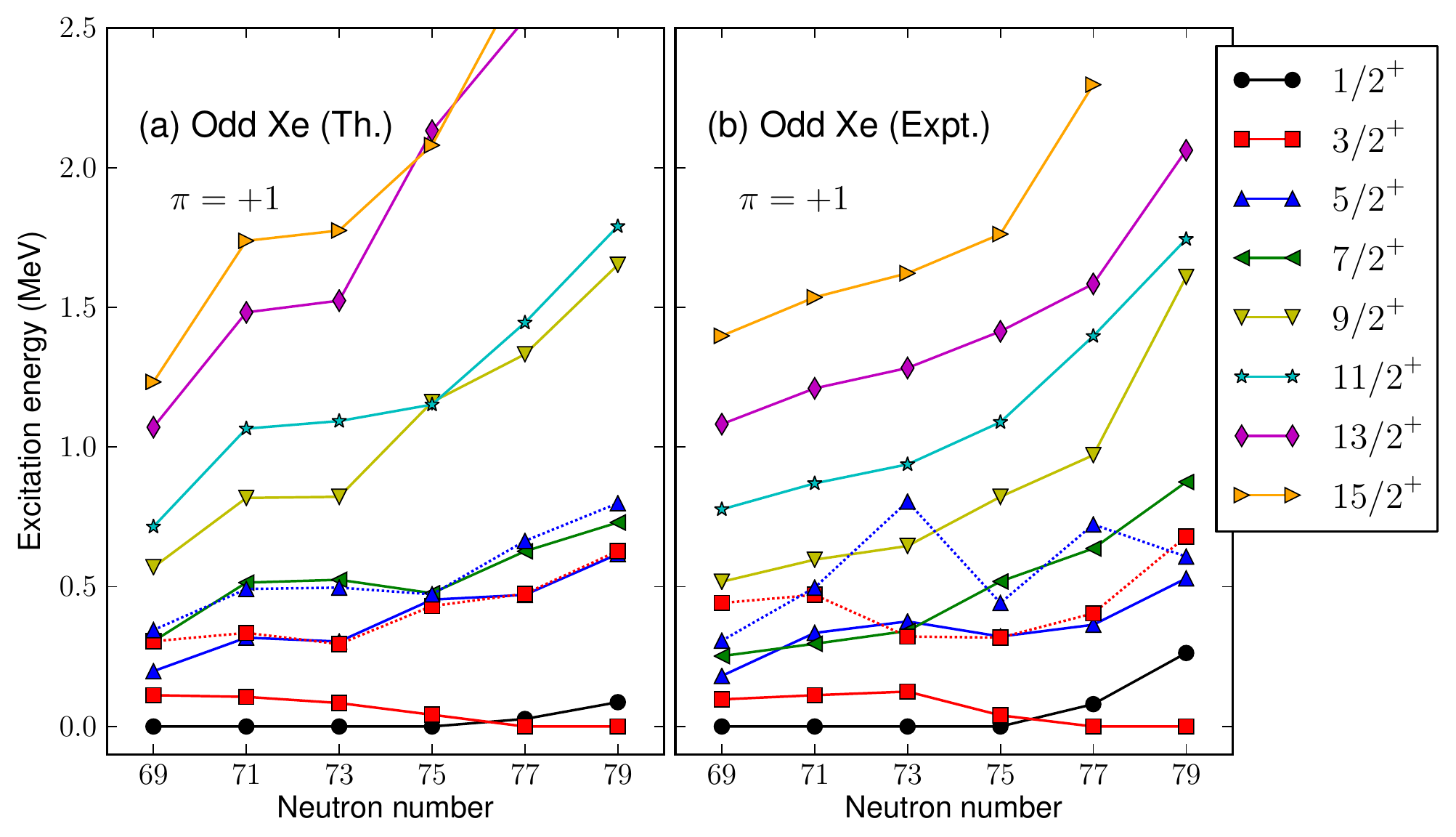}
\caption{(Color online) 
The calculated positive-parity low-energy
excitation spectra obtained 
for the odd-A  isotopes $^{123-133}$Xe are 
compared with the experimental data \cite{data}
taken from the ENSDF database. 
The  non-yrast ${3/2}^+_2$ and ${5/2}^+_2$ energy levels 
are connected by broken lines.}
\label{fig:level-oddxe}
\end{center}
\end{figure*}

\begin{figure*}[htb!]
\begin{center}
\includegraphics[width=0.7\linewidth]{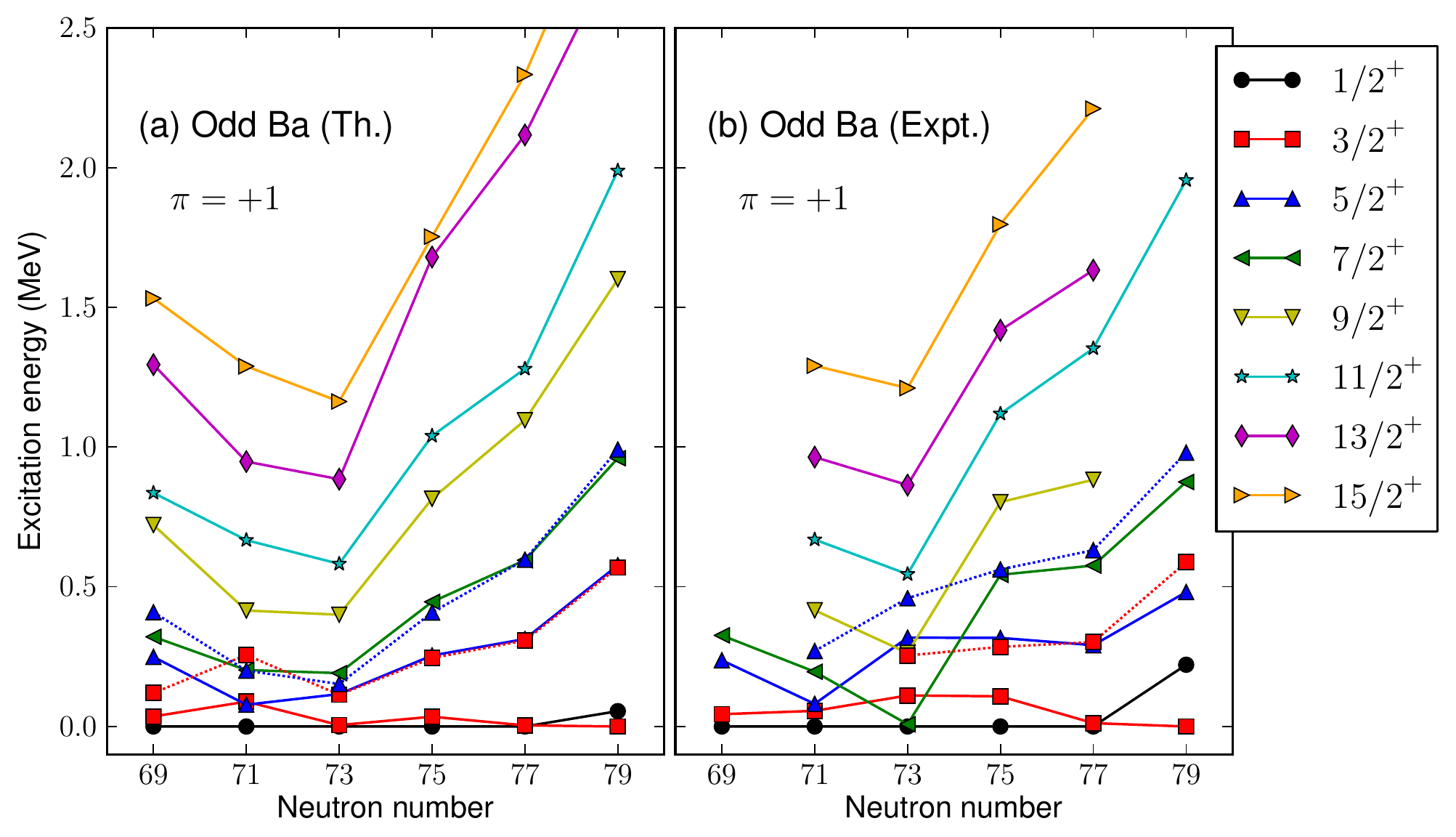}
\caption{(Color online) 
The same as in Fig.~\ref{fig:level-oddxe}, but for the odd-A  isotopes $^{125-135}$Ba.}
\label{fig:level-oddba}
\end{center}
\end{figure*}

\begin{figure*}[htb!]
\begin{center}
\includegraphics[width=0.7\linewidth]{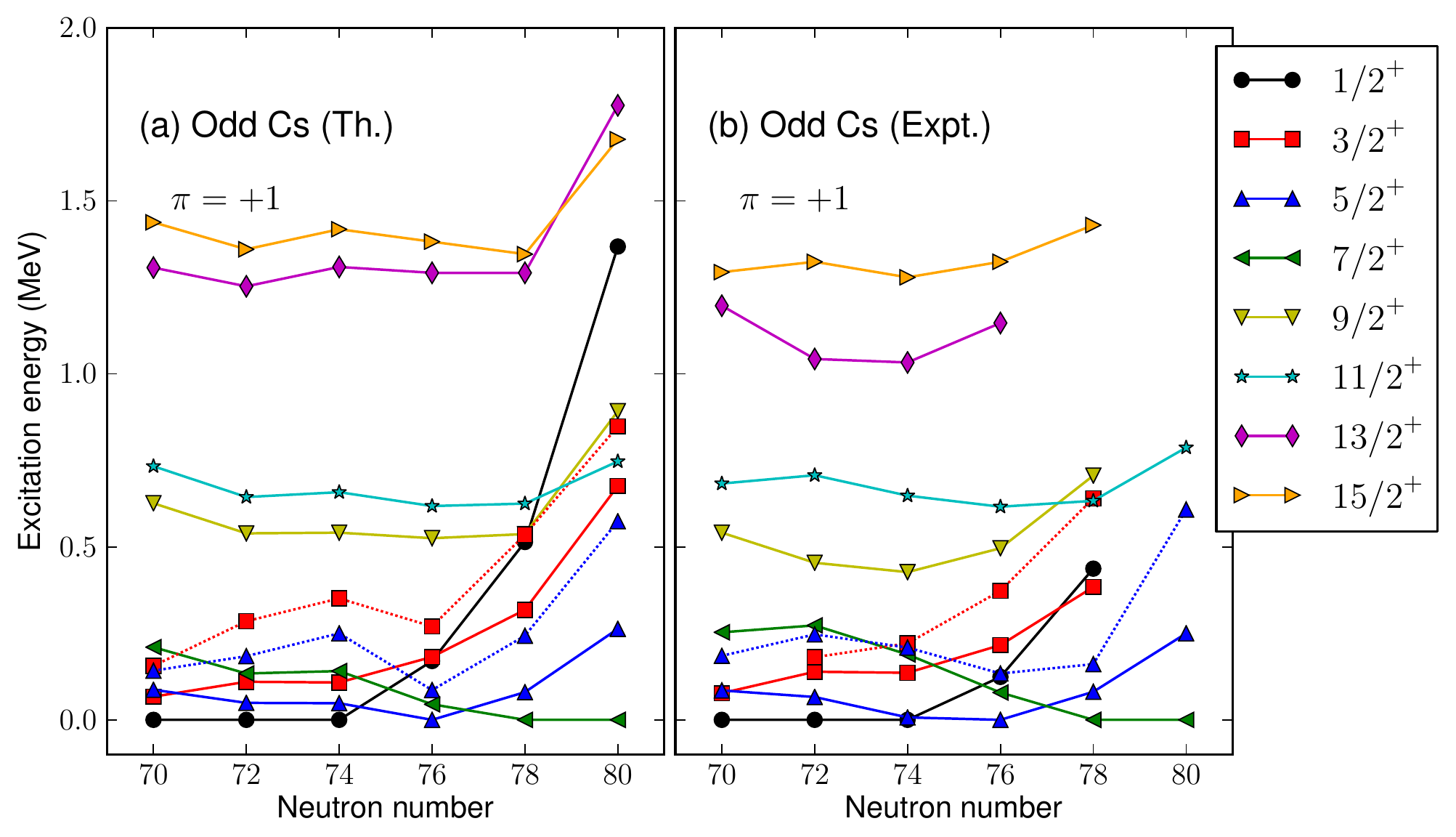}
\caption{(Color online) 
The same as in Fig.~\ref{fig:level-oddxe}, but for the odd-A  isotopes $^{125-135}$Cs.}
\label{fig:level-oddcs}
\end{center}
\end{figure*}

\begin{figure*}[htb!]
\begin{center}
\includegraphics[width=0.7\linewidth]{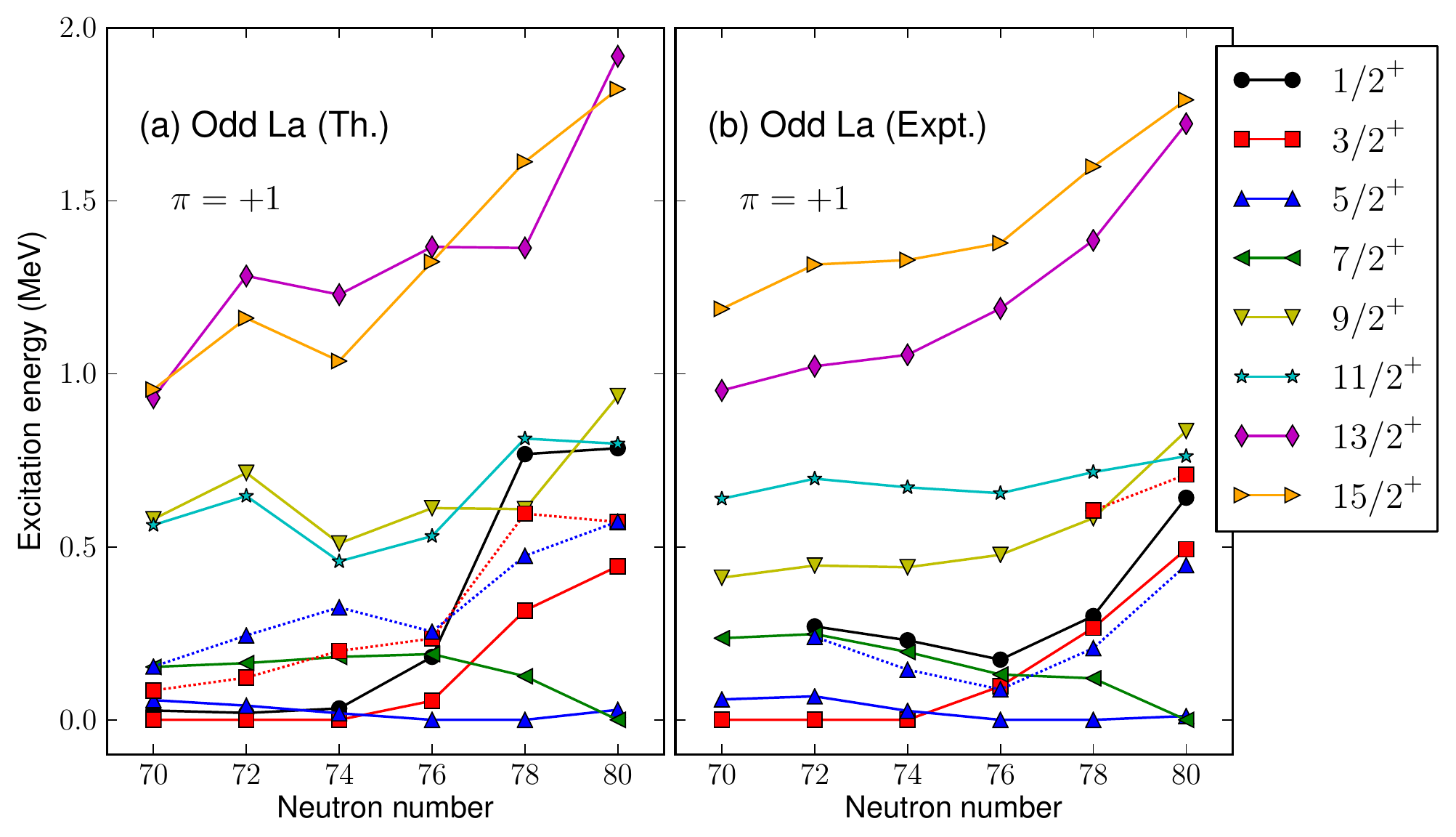}
\caption{(Color online) 
The same as in Fig.~\ref{fig:level-oddxe}, but for the odd-A  isotopes $^{127-137}$La.}
\label{fig:level-oddla}
\end{center}
\end{figure*}

\begin{figure*}[htb!]
\begin{center}
\includegraphics[width=\linewidth]{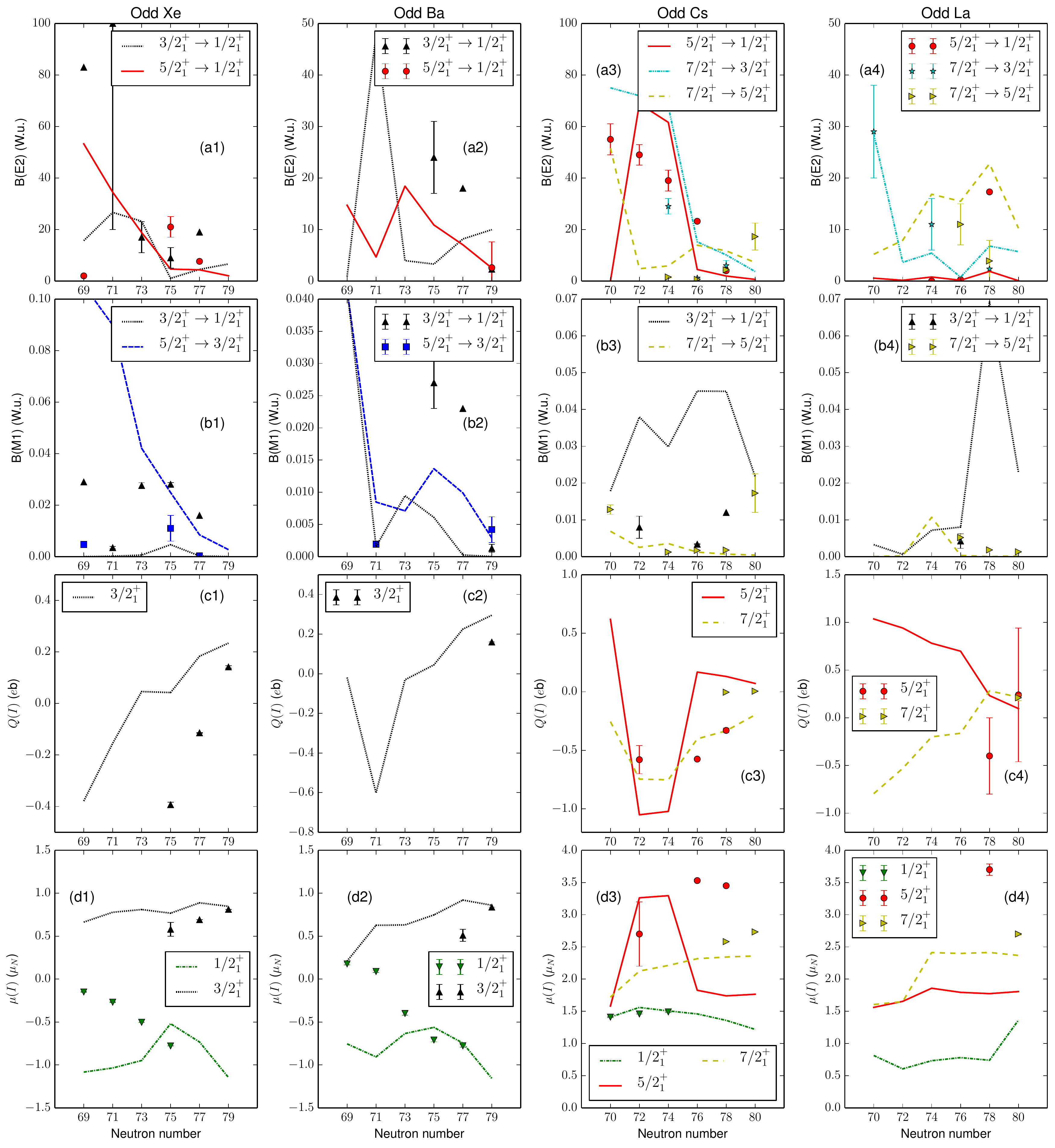}
\caption{(Color online) 
The $B$(E2) and $B$(M1)  transition rates (in Weisskopf units) as well as 
the 
electric quadrupole 
$Q(I)$ (in $e$b) and magnetic dipole $\mu(I)$ 
(in nuclear magneton $\mu_N$)
moments
corresponding to the lowest positive-parity states 
in the studied 
Xe (panels (a1--d1)), Ba (panels (a2--d2)), Cs (panels (a3--d3))
and La (panels (a4--d4)) nuclei are compared 
with the  
experimental data \cite{data}.
Those experimental $B$(E2) and $B$(M1) values 
without error bars are the lower limits. 
For the $B$(E2) rates, the theoretical and experimental 
values for the odd-N Xe and Ba (odd-Z Cs and La) nuclei 
are defined in panels (a1) and (a2) (panels (a3) and (a4)), 
respectively. 
The same rule applies to those other properties shown in the figure. 
}
\label{fig:em}
\end{center}
\end{figure*}

The positive-parity low-energy excitation spectra, obtained 
within the IBFM,  for odd-N Xe and  Ba as well as 
odd-Z Cs and La  nuclei are compared 
with the experimental spectra 
in Figs.~\ref{fig:level-oddxe}, \ref{fig:level-oddba}, \ref{fig:level-oddcs}, 
and \ref{fig:level-oddla}, respectively. As can be seen from 
the figures, the overall description of the empirical energy levels in each odd-A system
is very reasonable, in spite of having only used 
three fitted strength parameters to reproduce them.  
The evolution of some low-lying states can be associated with a shape transition.
For instance, in the  Xe isotopic chain 
(Fig.~\ref{fig:level-oddxe}) 
the ground state spin switches  from 
$I={1/2}^+$ at $N=75$ to
${3/2}^+$ at $N=77$. This correlates well
with the shape transitions observed in 
the neighboring even-even core nuclei
(see, Fig.~\ref{fig:level-even}). Perhaps, the 
most notable discrepancy between the calculated and experimental 
spectra is observed in the case of the ${1/2}^+_1$ state in La 
isotopes. The predicted energy levels   
are too low  for $N\leqslant 74$, as compared with their experimental counterparts.
However, in most of the La isotopes the experimental ${1/2}^+_1$ energy level 
has not been firmly established.

The $B$(E2) and $B$(M1)  transition rates as well as the 
electric quadrupole 
$Q(I)$ and magnetic dipole $\mu(I)$ moments
corresponding to the lowest  positive-parity states 
in all the considered odd-A 
Xe, Ba, Cs, and La nuclei are compared 
in Fig.~\ref{fig:em} with the available 
experimental data \cite{data}.
The electromagnetic transitions between the lowest-lying 
states tend to be stronger as the number of valence 
neutrons (holes) increases 
towards the middle of the major shell $N\approx 66$. 
Considerable differences between the computed 
and experimental  $B$(E2) and $B$(M1) values  are 
observed for the lightest Xe isotopes. In order 
to understand these deviations, it is useful to decompose  
the IBFM wave functions for the relevant 
states into the single-particle configurations involved.
For instance, the ${1/2}^+_1$, ${3/2}^+_1$, and ${5/2}^+_1$ states 
for $^{123}$Xe, where particularly large 
discrepancies are observed between the calculated 
and experimental $B$(E2) and $B$(M1) rates 
(see, panels (a1) and (b1)), can be expressed 
schematically in the following way: 
\begin{align}
\ket{\frac{1}{2}^+_1}&=\left[0.68\ket{\nu s_{1/2}}+0.28\ket{\nu d_{5/2}}+\ldots\right]\otimes\ket{^{124}\mathrm{Xe}} \nonumber\\
\ket{\frac{3}{2}^+_1}&=\left[0.67\ket{\nu d_{3/2}}+0.16\ket{\nu g_{7/2}}+\ldots\right]\otimes\ket{^{124}\mathrm{Xe}}\nonumber\\
\ket{\frac{5}{2}^+_1}&=\left[0.45\ket{\nu s_{1/2}}+0.49\ket{\nu d_{5/2}}+\ldots\right]\otimes\ket{^{124}\mathrm{Xe}},\nonumber \\
\end{align}
where the components with amplitudes smaller than 0.1 have been
omitted. The ${1/2}^+_1$ and ${5/2}^+_1$ states appear to be similar 
in structure, i.e., they are mainly 
made of the $3s_{1/2}$ and $2d_{5/2}$ single-neutron 
configurations. The  large overlap between the states 
leads to the strong $B$(E2) transition that follows 
the $\Delta I=2$ sequence of the weak coupling limit.
However, the ${1/2}^+_1$ and ${3/2}^+_1$ states have different 
structures leading to the small 
$B$(E2) and $B$(M1) transition rates computed between these states.
As for the electric quadrupole and magnetic dipole moments, shown in 
panels (c1,c2,c3,c4) and (d1,d2,d3,d4), the calculations 
reproduce reasonably well the experimental 
data, at least the correct sign, for most of the considered nuclei.

\section{$\beta$ decay\label{sec:beta}}

\subsection{Overall results}

\begin{figure}[htb!]
\begin{center}
\includegraphics[width=\linewidth]{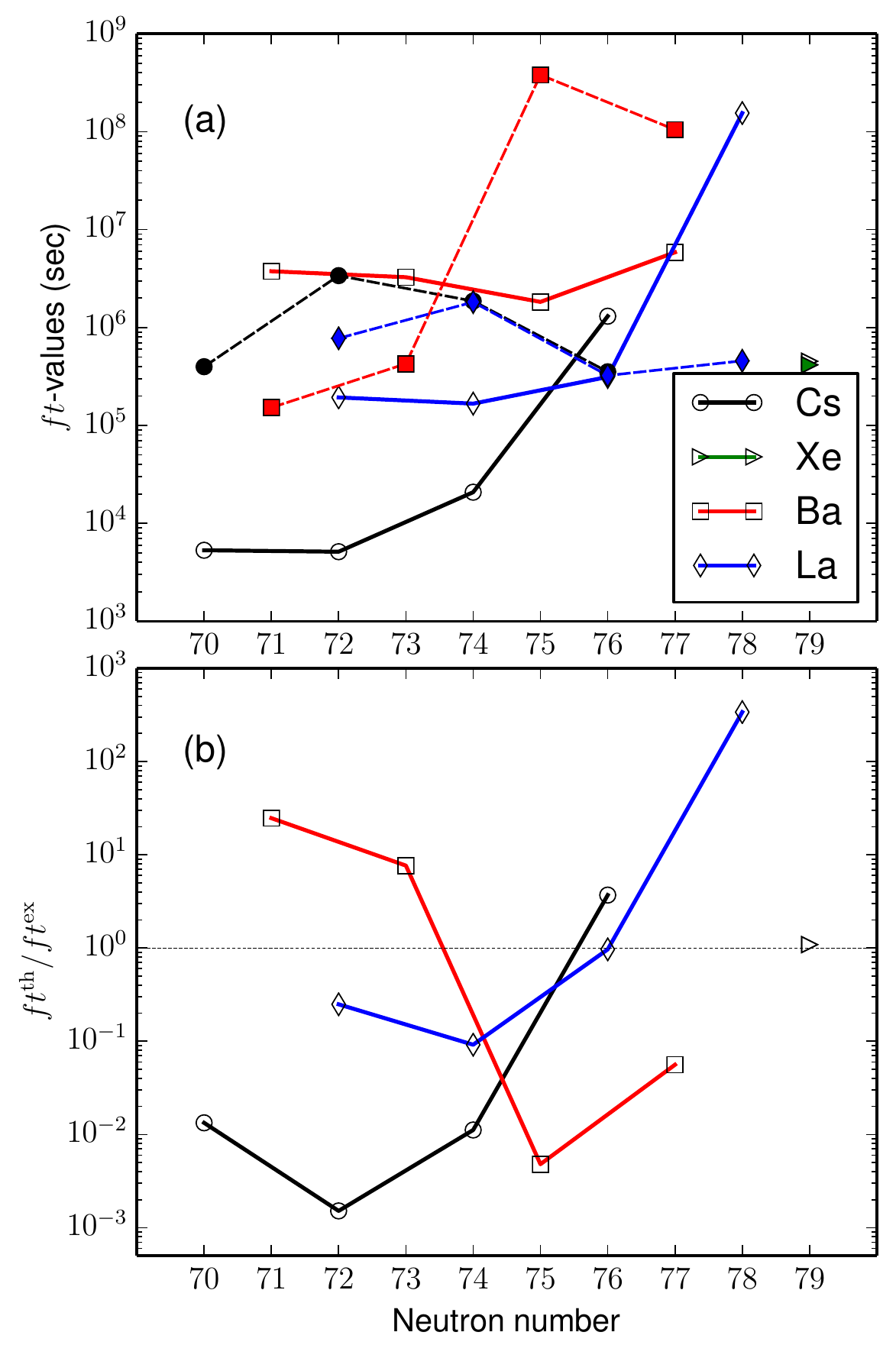}\\
\caption{(Color online) 
The $ft$-values corresponding to the 
$\beta^+$-decay of the odd-A nuclei $^{125-131}$Cs, 
$^{127-133}$Ba, and $^{129-135}$La
as well as the one corresponding to the 
$\beta^-$ decay of  $^{133}$Xe are plotted in panel (a).
The theoretical (experimental) values are represented 
by open (filled) symbols connected by solid (broken) lines.
The ratios $ft^\mathrm{th}/ft^\mathrm{ex}$
of the predicted to the experimental $ft$-values are depicted in panel (b).
}
\label{fig:ft}
\end{center}
\end{figure}

Having obtained a reasonable global description of the energies 
and transition properties in the even-even and odd-A nuclei, 
we now turn to the discussion of the $\beta$ decay. We have 
computed 
the $\log{ft}$ values of the 
$\beta^+$- and $\beta^-$-decays of those 
odd-A nuclei, for which the experimental 
data are available. We  will restrict our discussion to those 
$\beta$-decays where only GT and F transitions 
are involved, i.e., with $\Delta I=0,\pm 1$ and no parity change. 
Note, that 
most of the observed $\beta$-decays
in this region of the nuclear chart are of this type.
In panel (a) of Fig.~\ref{fig:ft}, we have plotted 
the $ft$-values corresponding to the 
$\beta^+$-decay of the odd-A nuclei $^{125-131}$Cs, 
$^{127-133}$Ba, and $^{129-135}$La
as well as the one corresponding to the 
$\beta^-$ decay of  $^{133}$Xe.
The transitions shown in the figure are those 
from the ground-state of parent nucleus to 
the lowest-energy state of daughter nucleus, 
for which experimental data are available.

The theoretical to experimental ratio for the $ft$-values, 
$ft^\mathrm{th}/ft^\mathrm{ex}$, are depicted in panel (b) of 
the same figure. The theoretical $ft$-values $ft^\mathrm{th}$
increase as the neutron shell closure $N=82$ is approached. 
In most of the cases, the computed 
$\beta$-decay $ft$-values 
underestimate the experimental ones, with the largest 
deviation for the Cs$\longrightarrow$Xe decays. 
In our calculation, the boson-core Hamiltonian, 
the single-particle energies and the occupation probabilities 
for the odd nucleon 
have been obtained via (constrained)  fully microscopic 
Gogny-D1M HFB calculations. In addition, 
no phenomenological parameter 
has been  introduced to compute the GT and 
F matrix elements. In view of these, 
the description of the observed $\log{ft}$ values by 
the present IBFM calculation is reasonably good. 
Let us also stress that no normalization factor has been introduced 
for the GT neither the  F matrix elements, as 
it is usually done, for example, in shell-model calculations.

\subsection{$\beta$ decays of odd-A Cs and Xe nuclei}

Let us now turn our attention to the detailed 
comparison between the computed and 
experimental $\log{ft}$ values
for the 
$\beta^\pm$ decays 
of the odd-A Cs and Xe (Table~\ref{tab:beta-csxe}).
In the case of the  odd-A Ba isotopes, the 
$\beta^+$ decays or electron-captures (EC)
are presented in Table~\ref{tab:beta-bacs}. The 
corresponding results for the
odd-A La nuclei 
can be found in 
Table~\ref{tab:beta-laba} and Table~\ref{tab:beta-laba2}. 
In each table, we have included the computed 
$\log{ft}$ values for transitions from an initial state 
to a selected set of final states (up to five 
states with the lowest energies for a given spin).
There are uncertainties in the experimental determination
of the spin of many nearly degenerate excited states that make
difficult to establish a correspondence with the states 
obtained in the calculation. In those cases, we use
a footnote in the experimental $\log{ft}$ value to indicate
the possible spins and parities. 
%In the case of odd-A nuclei, there are 
%many degenerated levels at high excitation energy.
%In order to avoid the complexity of the experimental
%$\log{ft}$ values for the transitions 
%to such states, in the tables 
%we have only shown the experimental
%data for the lowest two  degenerate levels
%(indicated in the footnotes of each table).

%\marginpar{Check this paragraph}

% ------------------- CHECKCHECK

\begin{table}[!htb]
\begin{center}
\caption{\label{tab:beta-csxe} 
The calculated and experimental $\log{ft}$ values 
for the $\beta^+$/EC (electron-capture) decay
of the odd-A Cs into Xe nuclei. Results are 
also included for the 
$\beta^-$ decay of $^{133}$Xe into $^{133}$Cs.
Experimental data are taken from Ref.~\cite{data}.}
\begin{ruledtabular}
 \begin{tabular}{cccc}
\multirow{2}{*}{Decay} &
\multirow{2}{*}{$I_\mathrm{i}\rightarrow  I_\mathrm{f}$} &  
\multicolumn{2}{c}{$\log{ft}$}\\
 \cline{3-4}
  & & Theory & Experiment \\
\hline
$^{125}$Cs$\rightarrow ^{125}$Xe
& ${1/2}^+_1\rightarrow {1/2}^+_1$ & 3.725 & $\approx$5.60 \\
& ${1/2}^+_1\rightarrow {1/2}^+_2$ & 4.608 & $\approx$5.53\footnotemark[1] \\
& ${1/2}^+_1\rightarrow {1/2}^+_3$ & 6.393 & $\approx$6.96\footnotemark[2]\\
%& *${1/2}^+_1\rightarrow {1/2}^+_4$ & 5.727 & $\approx$6.92 (1/2,3/2 at 1325)\\
%& *${1/2}^+_1\rightarrow {1/2}^+_5$ & 6.393 & $\approx$6.67 (1/2,3/2 at 1580)\\
& ${1/2}^+_1\rightarrow {3/2}^+_1$ & 5.871 & $\approx$6.76 \\
& ${1/2}^+_1\rightarrow {3/2}^+_2$ & 4.820 & $\approx$5.53\footnotemark[1] \\
& ${1/2}^+_1\rightarrow {3/2}^+_3$ & 4.771 & $\approx$6.38 \\
& ${1/2}^+_1\rightarrow {3/2}^+_4$ & 5.278 & $\approx$6.08 \\
& ${1/2}^+_1\rightarrow {3/2}^+_5$ & 5.045 & $\approx$6.96\footnotemark[2] \\
$^{127}$Cs$\rightarrow ^{127}$Xe
& ${1/2}^+_1\rightarrow {1/2}^+_1$ & 3.711 & 6.53(6) \\ % 
& ${1/2}^+_1\rightarrow {1/2}^+_2$ & 4.628 & 5.558(11) \\ % 4.161
& ${1/2}^+_1\rightarrow {1/2}^+_3$ & 5.151 & 7.305(18)\footnotemark[3]\\
& ${1/2}^+_1\rightarrow {1/2}^+_4$ & 5.984 & 8.02(3)\footnotemark[4]\\
%& ${1/2}^+_1\rightarrow {1/2}^+_5$ & 7.036 & 6.85(13)\footnotemark[5] at 1583\\
%& ${1/2}^+_1\rightarrow {1/2}^+_6$ & 5.009 & 6.73(3)\\
& ${1/2}^+_1\rightarrow {3/2}^+_1$ & 8.405 & 6.791(24) \\ % 5.727
& ${1/2}^+_1\rightarrow {3/2}^+_2$ & 4.426 & 7.574(20) \\ % 7.904
& ${1/2}^+_1\rightarrow {3/2}^+_3$ & 5.127 & 8.83(10) \\ % 4.776
& ${1/2}^+_1\rightarrow {3/2}^+_4$ & 5.563 & 6.306(12) \\ % 5.652
& ${1/2}^+_1\rightarrow {3/2}^+_5$ & 5.417 & 6.988(18) \\ % 6.637
%& ${1/2}^+_1\rightarrow {3/2}^+_6$ & 8.525 & 7.305(18)\footnotemark[3]at 1197 \\
%
$^{129}$Cs$\rightarrow ^{129}$Xe
& ${1/2}^+_1\rightarrow {1/2}^+_1$ & 4.318 & 6.27(6) \\
& ${1/2}^+_1\rightarrow {1/2}^+_2$ & 4.007 & 5.68(3) \\
& ${1/2}^+_1\rightarrow {1/2}^+_3$ & 4.201 & 6.80(4)\footnotemark[5] \\
& ${1/2}^+_1\rightarrow {3/2}^+_1$ & 6.728 & 7.3(2)\\
& ${1/2}^+_1\rightarrow {3/2}^+_2$ & 5.509 & 7.14(3)\\
& ${1/2}^+_1\rightarrow {3/2}^+_3$ & 4.694 & 6.50(3)\\
& ${1/2}^+_1\rightarrow {3/2}^+_4$ & 5.558 & 7.68(4) \\
& ${1/2}^+_1\rightarrow {3/2}^+_5$ & 6.403 & 6.80(4)\footnotemark[5]\\
$^{131}$Cs$\rightarrow ^{131}$Xe
& ${5/2}^+_1\rightarrow {3/2}^+_1$ & 6.116 & 5.548(14) \\
$^{133}$Xe$\rightarrow ^{133}$Cs
& ${3/2}^+_1\rightarrow {3/2}^+_1$ & 5.224 & 6.86(9) \\
& ${3/2}^+_1\rightarrow {5/2}^+_1$ & 5.656 & 5.619(12) \\
& ${3/2}^+_1\rightarrow {5/2}^+_2$ & 8.199 & 7.10(19)
\footnotetext[1]{${1/2}^{(+)},{3/2}^{(+)}$ level at 525 keV in $^{125}$Xe.}
\footnotetext[2]{${1/2},{3/2}$ level at 1312 keV in $^{125}$Xe.}
\footnotetext[3]{${1/2}^+,{3/2}^+$ level at 1197 keV in $^{127}$Xe}
\footnotetext[4]{${1/2},{3/2},{5/2}^+$ level at 1558 keV in $^{127}$Xe}
\footnotetext[5]{${1/2}^+$ or ${3/2}^+$ level at 946 keV in $^{129}$Xe}
 \end{tabular}
 \end{ruledtabular}
\end{center} 
\end{table}

For the $\beta$-decays of the odd-A Cs 
and Xe nuclei, the predicted 
$\log{ft}$ values  in Table~\ref{tab:beta-csxe} 
are systematically smaller than the experimental
ones. The discrepancy could be explained by analyzing the
dominant contributions to the GT and F 
transition matrix elements. As an example, let us consider
the decay of the ${1/2}^+_1$ 
ground state of $^{125}$Cs to the 
${1/2}^+_1$ ground state of $^{125}$Xe
for which the experimental 
$ft$-value is underestimated
by about a factor of 10$^2$ (see, Eq.~(\ref{eq:ft})).

The reduced GT transition matrix element 
for this decay is  
$\langle{1/2}^+_1||\hat{\cal O}^\mathrm{GT}||{1/2}^+_1\rangle=-1.194$,  
and the largest contributions come from  terms proportional to 
$s_\nu\times[a_{\nu s_{1/2}}^\dagger\times\tilde a_{\pi s_{1/2}}]^{(1)}$ 
and 
$[(a_{\nu s_{1/2}}^\dagger\times\tilde d_\nu)^{(j)}\times\tilde a_{\pi d_{5/2}}]^{(1)}$
(with $j$ being the intermediate angular momentum). The  
coefficients for these terms 
are $-0.900$ and $-0.306$, respectively. 
For the Fermi transition, 
$\langle{1/2}^+_1||\hat{\cal O}^\mathrm{F}||{1/2}^+_1\rangle=-0.145$, 
and the leading terms take the forms 
$s_\nu\times[a_{\nu s_{1/2}}^\dagger\times\tilde a_{\pi s_{1/2}}]^{(0)}$ 
and 
$s_\nu\times[a_{\nu d_{5/2}}^\dagger\times\tilde a_{\pi d_{5/2}}]^{(0)}$. 
Thus, the $3s_{1/2}$ and $2d_{5/2}$ neutron and proton 
single-particle configurations are dominant components 
in both the $B$(GT; ${1/2}^+_1\rightarrow{1/2}^+_1$) 
and 
$B$(F; ${1/2}^+_1\rightarrow{1/2}^+_1$) values for the 
$^{125}$Cs$\rightarrow^{125}$Xe decay. This agrees 
well , with the fact that the 
${1/2}^+_1$ IBFM ground state wave functions for 
the parent $^{125}$Cs and daugther $^{125}$Xe
systems mainly consist of the $3s_{1/2}$ 
and $2d_{5/2}$ single-particle configurations. In
particular, 31\% (67 \%) and 49\% (18\%) 
of the ${1/2}^+_1$ wave function in $^{125}$Cs ($^{125}$Xe) 
are accounted for by the $3s_{1/2}$ and $2d_{5/2}$ configurations, 
respectively. The similar  wave function contents 
for the  
parent and daughter nuclei could partly account for 
the too large $B$(GT) and $B$(F) values and, therefore, for the smaller 
$\beta$-decay $\log{ft}$ values as compared with the experimental data.

\subsection{$\beta$ decays of odd-A Ba nuclei}

\begin{table}[!htb]
\begin{center}
\caption{\label{tab:beta-bacs} 
The same as in Table~\ref{tab:beta-csxe}, but for 
the $\beta^+$/EC decays of the odd-A Ba into Cs nuclei.}
\begin{ruledtabular}
 \begin{tabular}{cccc}
\multirow{2}{*}{Decay} &
\multirow{2}{*}{$I_\mathrm{i}\rightarrow  I_\mathrm{f}$} & 
\multicolumn{2}{c}{$\log{ft}$}\\
 \cline{3-4}
  & & Theory & Experiment \\
\hline
%
%Ponchan
$^{127}$Ba$\rightarrow ^{127}$Cs
& ${1/2}^+_1\rightarrow {1/2}^+_1$ & 6.575 & 5.182(24) \\
& ${1/2}^+_1\rightarrow {1/2}^+_2$ & 5.357 & 7.25(15)\footnotemark[1]\\
& ${1/2}^+_1\rightarrow {1/2}^+_3$ & 5.354 & 7.25(9)\footnotemark[2]\\
& ${1/2}^+_1\rightarrow {3/2}^+_1$ & 6.736 & 6.81(11) \\
& ${1/2}^+_1\rightarrow {3/2}^+_2$ & 5.630 & 5.40(7) \\
& ${1/2}^+_1\rightarrow {3/2}^+_3$ & 5.629 & 7.25(15)\footnotemark[1]\\
& ${1/2}^+_1\rightarrow {3/2}^+_4$ & 6.732 & 7.25(9)\footnotemark[2]\\
$^{129}$Ba$\rightarrow ^{129}$Cs
& ${1/2}^+_1\rightarrow {1/2}^+_1$ & 6.514 & 5.63(3) \\
& ${1/2}^+_1\rightarrow {1/2}^+_2$ & 5.723 & 6.59(5)\footnotemark[3] \\
& ${1/2}^+_1\rightarrow {1/2}^+_3$ & 5.432 & 6.65(5)\footnotemark[4] \\
& ${1/2}^+_1\rightarrow {3/2}^+_1$ & 8.313 & 6.39(4) \\
& ${1/2}^+_1\rightarrow {3/2}^+_2$ & 6.483 & 5.91(3) \\
& ${1/2}^+_1\rightarrow {3/2}^+_3$ & 6.043 & 6.59(5)\footnotemark[3] \\
& ${1/2}^+_1\rightarrow {3/2}^+_4$ & 6.992 & 6.65(5)\footnotemark[4] \\
$^{131}$Ba$\rightarrow ^{131}$Cs
& ${1/2}^+_1\rightarrow {1/2}^+_1$ & 6.262 & 8.58(16) \\
& ${1/2}^+_1\rightarrow {1/2}^+_2$ & 5.892 & 6.633(8) \\
& ${1/2}^+_1\rightarrow {1/2}^+_3$ & 5.305 & 6.66(17)\footnotemark[5]\\
& ${1/2}^+_1\rightarrow {3/2}^+_1$ & 6.360 & 7.404(11) \\
& ${1/2}^+_1\rightarrow {3/2}^+_2$ & 6.729 & 7.305(9) \\
& ${1/2}^+_1\rightarrow {3/2}^+_3$ & 6.117 & 8.156(19) \\
& ${1/2}^+_1\rightarrow {3/2}^+_4$ & 7.256 & 8.505(11) \\
& ${1/2}^+_1\rightarrow {3/2}^+_5$ & 8.046 & 9.78(7)\footnotemark[6]\\
$^{133}$Ba$\rightarrow ^{133}$Cs
& ${1/2}^+_1\rightarrow {1/2}^+_1$ & 5.530 & 6.627(18) \\
& ${1/2}^+_1\rightarrow {3/2}^+_1$ & 6.768 & 8.020(15) 
\footnotetext[1]{${1/2},{3/2}$ level at 568 keV in $^{127}$Cs.}
\footnotetext[2]{${1/2},{3/2}$ level at 713 keV in $^{127}$Cs.}
\footnotetext[3]{$({1/2},{3/2})^+$ level at 554 keV in $^{129}$Cs.}
\footnotetext[4]{$({1/2},{3/2})^+$ level at 1165 keV in $^{129}$Cs.}
\footnotetext[5]{${1/2},{3/2}$ level at 1342 keV in $^{131}$Cs.}
\footnotetext[6]{$({3/2}^+,{3/2}^+)$ level at 920 keV in $^{131}$Cs.}
 \end{tabular}
 \end{ruledtabular}
\end{center} 
\end{table}

As seen from Table~\ref{tab:beta-bacs}, 
the $\log{ft}$ values predicted for the decays of 
the odd-A Ba nuclei are, in general, larger 
than those for the decays of odd-A Cs (considered 
in Table~\ref{tab:beta-csxe}). 
The overall description of the experimental $\log{ft}$ 
values in the cases of the Ba$\longrightarrow$Cs 
decays is, therefore, slightly better than 
for the Cs$\longrightarrow$Xe decays. 

We observe that, similar to the cases of the 
odd-A Cs nuclei, 
the predicted $\log{ft}$ values of the odd-A Ba are 
calculated to be systematically smaller
than the empirical values. 
There are, however, examples of  $\beta$ decays of the 
odd-A Ba systems, for which the theory overestimates the experiment.
The largest deviation 
occurs, for example, in the case of the  
$^{129}$Ba$({1/2}^+_1)\rightarrow^{129}$Cs$({3/2}^+_1)$ decay.
In particular, the computed $\log{ft}$ is  
a factor of 1.3 larger than the experimental one.
Here, only the GT transition is involved.
The reduced GT  matrix element  
is found to be as small as 0.0061, due to the cancellation of the many small components
that make the matrix element.
The largest contributions come from  terms of the type 
$s_\nu^\dagger s_\pi [[\tilde d_\nu\times a_{\nu s_{1/2}}^\dagger]^{(3/2)}\times\tilde a_{\pi d_{5/2}}]^{(1)}$
and 
$s_\nu^\dagger s_\pi [[\tilde d_\nu\times a_{\nu s_{1/2}}^\dagger]^{(5/2)}\times\tilde a_{\pi d_{5/2}}]^{(1)}$. 
However, their coefficients are 0.0447 and $-0.0334$, respectively.
As a result, a small $B$(GT) is obtained and this leads 
to a too large $\log{ft}$ value as compared with the experiment.
The  ${1/2}^+_1$ ground state IBFM wave function of 
the parent nucleus 
$^{129}$Ba is mainly 
made of the 
$\nu s_{1/2}$ (36 \%) and $\nu d_{3/2}$ (47 \%) 
single-neutron configurations  while, 
the  ${3/2}^+_1$ wave function corresponding to the  
daughter system 
$^{129}$Cs is dominated by the $\pi g_{7/2}$ (41\%) and 
$\pi d_{3/2}$ (31 \%) configurations. This 
difference in the parent and daughter states may partly 
account for the small 
GT strength.

\subsection{$\beta$ decays of odd-A La nuclei}

\begin{table}[!htb]
\begin{center}
\caption{\label{tab:beta-laba} 
The same as in Table~\ref{tab:beta-csxe}, but for the 
$\beta^+$/EC decays of the 
odd-A $^{129,131}$La nuclei. }
\begin{ruledtabular}
 \begin{tabular}{cccc}
\multirow{2}{*}{Decay} &
\multirow{2}{*}{$I_\mathrm{i}\rightarrow  I_\mathrm{f}$} &  
\multicolumn{2}{c}{$\log{ft}$}\\
 \cline{3-4}
  & & Theory & Experiment \\
\hline
$^{129}$La$\rightarrow^{129}$Ba
%Ponchan
& ${3/2}^+_1\rightarrow {1/2}^+_1$ & 5.286 & 5.89(8) \\
& ${3/2}^+_1\rightarrow {1/2}^+_2$ & 6.559 & 5.55(2)\\
& ${3/2}^+_1\rightarrow {1/2}^+_3$ & 4.871 & 7.2(1) \\
& ${3/2}^+_1\rightarrow {1/2}^+_4$ & 5.477 & 6.83(7) \\
& ${3/2}^+_1\rightarrow {1/2}^+_5$ & 9.064 & 6.62(5) \\
%& ${3/2}^+_1\rightarrow {1/2}^+_6$ & 7.382 & 6.60(5) \\
& ${3/2}^+_1\rightarrow {3/2}^+_1$ & 4.713 & 5.90(5) \\
& ${3/2}^+_1\rightarrow {3/2}^+_2$ & 6.058 & 6.23(5)\\
& ${3/2}^+_1\rightarrow {3/2}^+_3$ & 6.856 & 5.59(3)\\
& ${3/2}^+_1\rightarrow {3/2}^+_4$ & 6.916 & 6.45(4)\footnotemark[1]\\
& ${3/2}^+_1\rightarrow {3/2}^+_5$ & 4.502 & 6.06(3)\footnotemark[2]\\
%& ${3/2}^+_1\rightarrow {3/2}^+_6$ & 5.004 & 6.31(4) at 889\\
%& ${3/2}^+_1\rightarrow {3/2}^+_7$ & 6.266 & 7.3(1)\\
& ${3/2}^+_1\rightarrow {5/2}^+_1$ & 6.991 & 6.60(5) \\
& ${3/2}^+_1\rightarrow {5/2}^+_2$ & 6.149 & 7.3(1) \\
& ${3/2}^+_1\rightarrow {5/2}^+_3$ & 7.436 & 6.45(4)\footnotemark[1]\\
& ${3/2}^+_1\rightarrow {5/2}^+_4$ & 7.978 & 6.50(4)\\
& ${3/2}^+_1\rightarrow {5/2}^+_5$ & 5.426 & 6.06(3)\footnotemark[2]\\
%& ${3/2}^+_1\rightarrow {5/2}^+_6$ & 5.773 & 7.1(2)\\
%
$^{131}$La$\rightarrow ^{131}$Ba
& ${3/2}^+_1\rightarrow {1/2}^+_1$ & 5.223 & 6.26(9)\\
& ${3/2}^+_1\rightarrow {1/2}^+_2$ & 6.710& 5.82(3)\\
& ${3/2}^+_1\rightarrow {1/2}^+_3$ & 5.385 & 6.56(4)\footnotemark[3] \\
& ${3/2}^+_1\rightarrow {1/2}^+_4$ & 5.114 & 6.43(3)\footnotemark[4] \\
& ${3/2}^+_1\rightarrow {3/2}^+_1$ & 4.676 & 6.25(5) \\
& ${3/2}^+_1\rightarrow {3/2}^+_2$ & 5.857 & 6.34(4)\\
& ${3/2}^+_1\rightarrow {3/2}^+_3$ & 7.121 & 5.58(3)\\
& ${3/2}^+_1\rightarrow {3/2}^+_4$ & 5.345 & 6.18(3)\footnotemark[5]\\
& ${3/2}^+_1\rightarrow {3/2}^+_5$ & 6.077 & 6.56(4)\footnotemark[3]\\
& ${3/2}^+_1\rightarrow {5/2}^+_1$ & 7.193 & 6.85(5)\\
& ${3/2}^+_1\rightarrow {5/2}^+_2$ & 5.917 & 6.18(3)\footnotemark[5]\\
& ${3/2}^+_1\rightarrow {5/2}^+_3$ & 5.601 & 6.56(4)\footnotemark[3]
%& *${3/2}^+_1\rightarrow {5/2}^+_4$ & 6.886 & 6.43(3)\footnotemark[4]
%
\footnotetext[1]{$({3/2}^{+},{5/2}^{+})$ level at 618 keV in $^{129}$Ba.}
\footnotetext[2]{$({3/2}, {5/2})^+$ level at 712 keV in $^{129}$Ba.}
\footnotetext[3]{${1/2}^+, {3/2}^+, {5/2}^+$ level at 719 keV in $^{131}$Ba.}
\footnotetext[4]{${1/2}^+, {3/2}^+, {5/2}^+$ level at 879 keV in $^{131}$Ba.}
\footnotetext[5]{${3/2}^+, {5/2}^+$ level at 562 keV in $^{131}$Ba.}
 \end{tabular}
 \end{ruledtabular}
\end{center} 
\end{table}

\begin{table}[!htb]
\begin{center}
\caption{\label{tab:beta-laba2} 
The same as in Table~\ref{tab:beta-csxe}, but for the 
$\beta^+$/EC decays of the odd-A $^{133,135}$La nuclei.}
\begin{ruledtabular}
 \begin{tabular}{cccc}
\multirow{2}{*}{Decay} &
\multirow{2}{*}{$I_\mathrm{i}\rightarrow  I_\mathrm{f}$} &  
\multicolumn{2}{c}{$\log{ft}$}\\
 \cline{3-4}
  & & Theory & Experiment \\
\hline
$^{133}$La$\rightarrow ^{133}$Ba
& ${5/2}^+_1\rightarrow {3/2}^+_1$ & 5.495 & 5.51(4)\\
& ${5/2}^+_1\rightarrow {3/2}^+_2$ & 5.420 & 6.97(3)\\
& ${5/2}^+_1\rightarrow {3/2}^+_3$ & 7.293 & 6.89(3)\\
& ${5/2}^+_1\rightarrow {3/2}^+_4$ & 7.635 & 7.77(4)\\
& ${5/2}^+_1\rightarrow {3/2}^+_5$ & 9.694 & 8.48(9)\footnotemark[1]\\
& ${5/2}^+_1\rightarrow {5/2}^+_1$ & 5.621 & 7.26(5)\\
& ${5/2}^+_1\rightarrow {5/2}^+_2$ & 5.375 & 7.13(3)\\
& ${5/2}^+_1\rightarrow {5/2}^+_3$ & 5.192 & 7.80(6)\footnotemark[2]\\
& ${5/2}^+_1\rightarrow {5/2}^+_4$ & 5.490 & 6.71(4)\\
& ${5/2}^+_1\rightarrow {5/2}^+_5$ & 5.718 & 7.24(4)\footnotemark[3]\\
%& ${5/2}^+_1\rightarrow {5/2}^+_6$ & 7.747 & 7.42(5) at 1212\\
%& ${5/2}^+_1\rightarrow {5/2}^+_7$ & 4.997 & 7.61(5)\\
& ${5/2}^+_1\rightarrow {7/2}^+_1$ & 5.770 & 7.51(4)\\
& ${5/2}^+_1\rightarrow {7/2}^+_2$ & 5.671 & 8.13(5)\\
& ${5/2}^+_1\rightarrow {7/2}^+_3$ & 7.105 & 7.24(4)\footnotemark[3]\\
& ${5/2}^+_1\rightarrow {7/2}^+_4$ & 5.498 & 7.36(5)\\
$^{135}$La$\rightarrow ^{135}$Ba
& ${5/2}^+_1\rightarrow {3/2}^+_1$ & 8.190 & 5.66(1)\\
& ${5/2}^+_1\rightarrow {3/2}^+_2$ & 5.577 & 7.88(7)\\
& ${5/2}^+_1\rightarrow {3/2}^+_3$ & 8.599 & 7.75(8)\\
& ${5/2}^+_1\rightarrow {3/2}^+_4$ & 5.634 & 8.25(9)\footnotemark[4]\\
& ${5/2}^+_1\rightarrow {5/2}^+_1$ & 4.961 & 7.01(7)\\
& ${5/2}^+_1\rightarrow {5/2}^+_2$ & 5.237 & 8.25(9)\footnotemark[4]\\
& ${5/2}^+_1\rightarrow {7/2}^+_1$ & 5.531 & 7.22(9)
\footnotetext[1]{${3/2}, {5/2}^+$ level at 1528 keV in $^{133}$Ba.}
\footnotetext[2]{${3/2}^+, {5/2}^+$ level at 676 keV in $^{133}$Ba.}
\footnotetext[3]{${3/2}^+, {5/2}^+, {7/2}^+$ level at 1112 keV in $^{133}$Ba.}
\footnotetext[4]{${3/2}^+, {5/2}^+$ level at 980 keV in $^{135}$Ba.}
 \end{tabular}
 \end{ruledtabular}
\end{center} 
\end{table}

From Table~\ref{tab:beta-laba} 
and Table~\ref{tab:beta-laba2} one sees that the 
computed $\log{ft}$ values for the $\beta^+$ decays
La$\longrightarrow$Ba are larger than the ones 
obtained for the Cs$\longleftrightarrow$Xe 
and Ba$\longrightarrow$Cs decays 
(see, Table~\ref{tab:beta-csxe} 
and Table~\ref{tab:beta-bacs}). However, the 
$\log{ft}$ values for some of the La$\longrightarrow$Ba 
transitions are still too small as compared 
to the experimental ones \cite{data}. This 
mainly occurs (see, Table~\ref{tab:beta-laba})
for the $\beta$ decay with $\Delta I=0$, where the $B$(GT) 
as well as the $B$(F) transition strengths 
are too large. 
A typical example is the decay from the 
${5/2}^+_1$ ground state 
of $^{135}$La to the ${5/2}^+_1$ state of $^{135}$Ba. 
For this transition, the theoretical  result underestimates the 
experimental ${ft}$-value by a factor of $\approx 10^2$ 
(see, Table~\ref{tab:beta-laba2} and Eq.~(\ref{eq:ft})).
The dominant contribution to the 
GT transition strength comes from the  term 
$s_\nu[a_{\nu d_{3/2}}^\dagger\times\tilde a_{\pi d_{5/2}}]^{(1)}$ 
with a coefficient in front of 0.414. Here, the 
$\pi d_{5/2}$ 
and $\nu d_{3/2}$ configurations play the dominant role 
for the parent and daughter nuclei, respectively. 
The ${5/2}^+_1$ wave function of 
$^{135}$La is  mainly based on the $\pi g_{7/2}$ configuration (89\%)
 while for $^{135}$Ba the 
${5/2}^+_1$ excited state   is mainly based
on the 
$\nu d_{3/2}$ configuration (79\%) with a 
 $\nu s_{1/2}$ 
component  (19\%). Therefore, it is tempting to 
interpret the unexpectedly large 
$B$(GT) value solely in terms of the relevant wave function contents. 
It must be kept in mind, however, that the $B$(GT), as well as the $B$(F), transition strengths
also depend upon the
coefficients for the one-particle transfer operators 
(see, Eqs.~(\ref{eq:creation1}--\ref{eq:annihilation2}))  and those
depend on other factors 
such as the occupation probabilities $v^2_j$ for the odd particle.
In this particular example, the relevant coefficient is 
$\eta_{\nu{3/2},\pi{5/2}}\theta_{\nu{3/2}}\zeta_{\pi{5/2}}$ 
(see Eqs.~(\ref{eq:zeta1}), (\ref{eq:theta1}), and (\ref{eq:eta}).

For those transitions, where the $B$(GT) 
and $B$(F) values are found 
to be too small, i.e., the resulting $\log{ft}$ values 
are too large, cancellation seems to occur 
to a large extent between different components 
of the operators. 
For instance, for the transition 
$^{135}$La$({5/2}^+_1)\longrightarrow^{135}$Ba$({3/2}^+_1)$, 
the largest terms in the GT matrix element 
turn out to be 
$s_\nu s_\pi [a^\dagger_{\nu d_{3/2}}\times(d_\pi^\dagger\times\tilde a_{\pi g_{7/2}})^{5/2}]^{(1)}$, 
$s_\nu s_\pi [a^\dagger_{\nu d_{3/2}}\times(d_\pi^\dagger\times\tilde a_{\pi d_{3/2}})^{5/2}]^{(1)}$, 
and 
$s_\nu(a^\dagger_{\nu d_{3/2}}\times\tilde a_{\pi d_{5/2}})^{(1)}$. 
The corresponding coefficients  ($-0.0562$, 
$-0.0246$, and $0.0979$)  
almost cancel each other. 
The IBFM wave functions for the 
${5/2}^+_1$ and ${3/2}^+_1$ 
states of  $^{135}$La and  
$^{135}$Ba  are mainly made  of 
the $\pi g_{7/2}$ and $\nu d_{3/2}$ configurations.
Once more, the $\log{ft}$ value for this transition is not 
completely accounted for by simply looking at the 
compositions of the wave functions. 
As we have already noted, the coefficients for the 
fermion transfer operators are determined by 
various factors such as the occupation probabilities 
$v^2_j$ and boson-core wave functions. In turn, those  
factors are determined 
microscopically from the Gogny-D1M mean-field 
results.

\subsection{Sensitivity to the IBFM parameters}

\begin{figure}[htb!]
\begin{center}
\includegraphics[width=\linewidth]{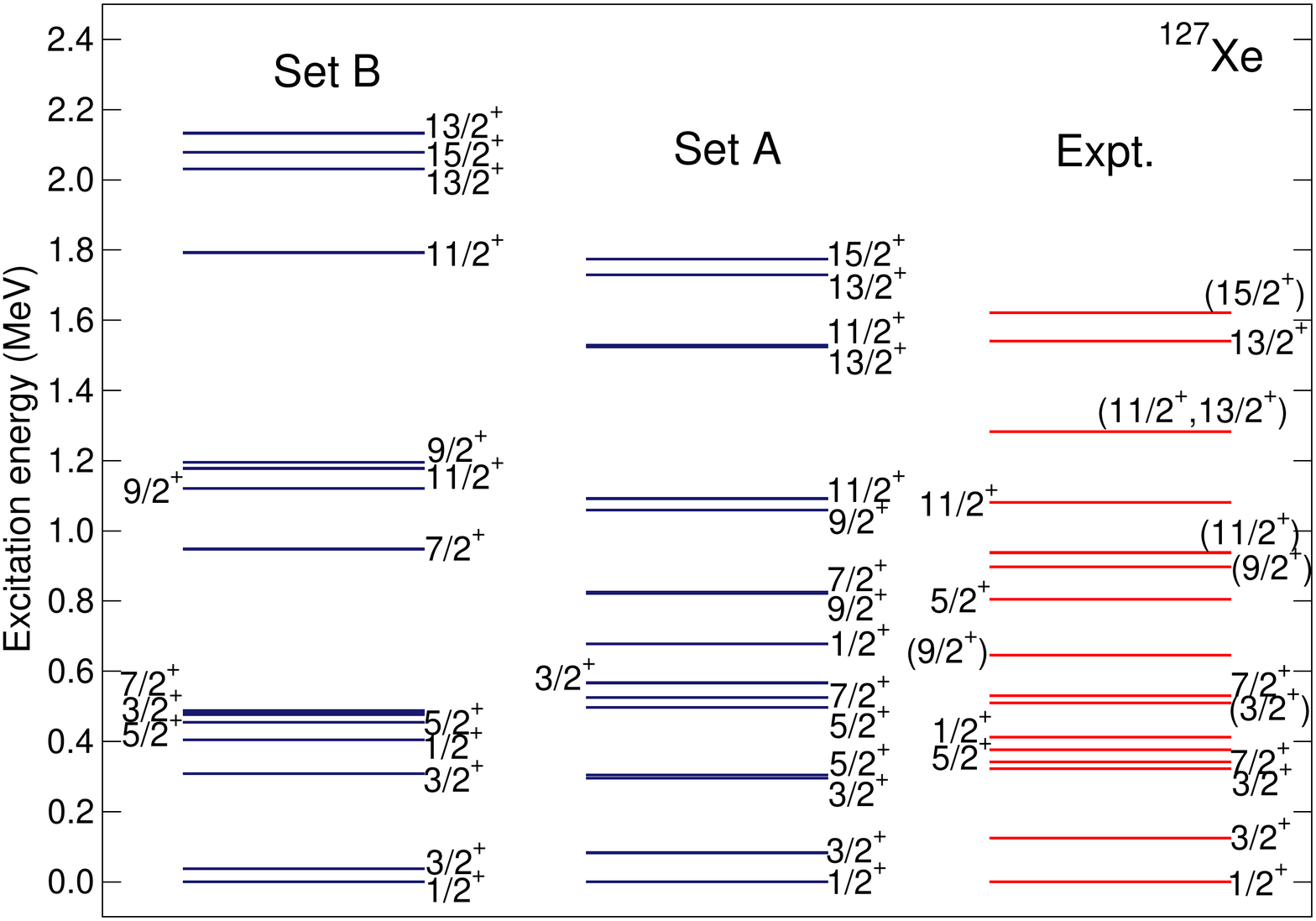}\\
\caption{(Color online) 
Level schemes for $^{127}$Xe. 
Results of  two independent IBFM calculations, employing different sets of
boson-fermion strengths, are compared with each 
other and with the experimental energy spectrum. The
original strength parameters, shown in Table~\ref{tab:ibfm2para}, are
denoted as {\it set A} while the modified strengths 
are denoted as {\it set B}. For more details, see the main text.
}
\label{fig:comp-xe127}
\end{center}
\end{figure}

As already mentioned in previous sections, the 
predicted $B$(GT) and $B$(F) values 
appear to be somewhat sensitive to the details 
of the IBFM wave functions for the parent and 
daughter nuclei. In what follows we study
the dependence of the computed 
$\beta$-decay 
$\log{ft}$ values on the 
strength parameters of 
the IBFM.

As an example, let us consider the $^{127}$Cs$\longrightarrow^{127}$Xe 
decay. We have performed additional IBFM calculations for 
$^{127}$Xe with the 
strengths 
$\Gamma_\nu=1.6$ MeV, $\Lambda_\nu=2.0$ MeV 
and $A_\nu=-0.0$ MeV. We will refer to this set of parameters 
as {\it set B} while the original 
strengths, shown in Table~\ref{tab:ibfm2para}, will 
be denoted as {\it set A}. 
In Fig.~\ref{fig:comp-xe127}, we have compared the energy spectra 
resulting from those two independent IBFM calculations 
with the experimental one. 
Both, the {\it set B} and {\it set A} IBFM calculations 
reproduce the experimental excitation energies of states in the vicinity of 
the ground state with the same level of accuracy.

However, at higher excitation energy, the values obtained with {\it set B} are much higher 
in energy than those with {\it set A} mostly for levels with spin $I\geqslant{9/2}$).
The description of the corresponding E2 and M1 transitions 
and moments is worse with  {\it set B}. However, 
the $\log{ft}$ values in the case of {\it set B}  
are improved with respect to the ones obtained with {\it set A}. 
For instance with {\it set B} we have 
obtained $\log{ft}$ = 4.486 for the 
${1/2}^+_1\rightarrow{1/2}^+_1$ transition from $^{127}$Cs to $^{127}$Xe.
This is closer to the experimental result $\log{ft}$ = 6.53 $\pm$ 0.06
than the value $\log{ft}$ = 3.711 obtained with {\it set A}.

The difference in the predicted $\log{ft}$ values mainly comes from the 
corresponding 
$B$(GT) rates. We have obtained 
the reduced GT matrix element $\langle{1/2}^+_1||{\cal{\hat O^\mathrm{GT}}}||{1/2}^+_1\rangle=-1.216$ 
with {\it set A}. In this case, the largest contribution comes from the term 
$s_\nu\times [a_{\nu s_{1/2}}^\dagger\times\tilde a_{\pi s_{1/2}}]^{(1)}$ 
with the coefficient of $-0.837$. 
The contributions from the $\nu d_{3/2}\rightarrow\pi d_{5/2}$ and 
 $\nu d_{5/2}\rightarrow\pi d_{5/2}$ terms 
 are non-negligible but enter with opposite signs and therefore cancel each other.
On the other hand, with {\it set B}, we have 
obtained 
$\langle{1/2}^+_1||{\cal{\hat O^\mathrm{GT}}}||{1/2}^+_1\rangle=0.448$
and the dominant component is the term 
$s_\nu\times [a_{\nu s_{1/2}}^\dagger\times\tilde a_{\pi s_{1/2}}]^{(1)}$ 
with the coefficient $0.520$.

The ${1/2}^+_1$ ground state for $^{127}$Xe reads 
\begin{align}
\ket{{\frac{1}{2}}^+_1}=
&\large(0.29\ket{\nu s_{1/2}}+0.55\ket{\nu d_{3/2}}\nonumber \\
&+0.12\ket{\nu d_{5/2}}+0.04\ket{\nu g_{7/2}}\large)\otimes\ket{^{128}\mathrm{Xe}}. 
\end{align}
with {\it set A} while with {\it set B} it takes the form 
\begin{align}
\ket{{\frac{1}{2}}^+_1}=
&\large(0.62\ket{\nu s_{1/2}}+0.19\ket{\nu d_{3/2}}\nonumber \\
&+0.16\ket{\nu d_{5/2}}+0.02\ket{\nu g_{7/2}}\large)\otimes\ket{^{128}\mathrm{Xe}}.
\end{align}
The two wave functions mainly differ  in the amplitudes of the 
$\nu s_{1/2}$ and $\nu d_{3/2}$ configurations.

The $2d_{3/2}$ single-particle orbital 
is the lowest in energy among the ones employed  
for $^{127}$Xe  
(see, Fig.~1 of Ref.~\cite{nomura2017odd-3}). 
Of the three boson-fermion terms, the exchange term 
is particularly important in mixing different single-particle 
configurations and the most significant difference, between 
the parameters of  {\it set A} and {\it set B}  is 
perhaps the larger exchange  strength 
$\Lambda_\nu$ used for {\it set B}. Due to this, the 
mixing of the $\nu s_{1/2}$ single-particle 
components into the ${1/2}^+_1$ ground state of $^{127}$Xe
is stronger.
On the other hand, the ${1/2}^+_1$ ground state of the 
parent nucleus $^{127}$Cs can be decomposed as follows:
\begin{align}
\ket{{\frac{1}{2}}^+_1}=
&\large(0.24\ket{\pi s_{1/2}}+0.08\ket{\pi d_{3/2}}\nonumber \\
&+0.66\ket{\pi d_{5/2}}+0.02\ket{\pi g_{7/2}}\large)\otimes\ket{^{126}\mathrm{Xe}}. 
\end{align}
which has a similar mixing amplitude for the 
$s_{1/2}$ single-proton configuration to the 
${1/2}^+_1$ wave function for the daughter nucleus.

% ----------------------------------------------------------------------

\section{Summary and concluding remarks\label{sec:summary}}

% ----------------------------------------------------------------------

In this paper, we have presented a consistent 
description of the low-energy 
excitation spectra and $\beta$ decay of odd-A nuclei 
within the IBFM based on input from realistic mean field calculations. 
The $(\beta,\gamma)$ potential energy surfaces for 
even-even nuclei, the spherical single particle 
energies and the occupation probabilities for the neighboring 
odd-A nuclei have been computed  microscopically 
within the  constrained HFB
scheme based on the Gogny-D1M EDF.
Those quantities are used as microscopic input to 
access the spectroscopic properties in odd-A nuclei 
within the IBFM. 
Only the three coupling constants for the 
boson-fermion interaction terms 
have been fitted to experimental data  as to
reproduce reasonably well the low-lying 
energy levels in each of the studied odd-A systems.
Having the IBFM wave functions for the parent and 
daughter nuclei, 
the GT and F transition strengths 
have been computed without any additional 
phenomenological parameter.

The low-lying  positive-parity excitation spectra 
are reproduced reasonably well
for the even-even 
Xe and Ba, the neighboring 
odd-N Xe and Ba as well as the odd-Z Cs and La nuclei. 
The isotopic dependence 
of the energy levels in the studied 
odd-A nuclei,  and in the corresponding even-even 
Xe and Ba cores,  points to an evolution 
from prolate  to $\gamma$-soft and to nearly spherical shapes. 
Electromagnetic properties, such as the $B$(E2) transition
rates and magnetic dipole moments, are well reproduced. 
%As in earlier $\beta$-decay IBFM studies, 
We have obtained $\beta$-decay $\log{ft}$ values that are systematically 
smaller than the experimental ones. 
Those $\log{ft}$ values mainly depend 
on the details of the IBFM wave functions for the 
parent and daughter nuclei and, in many cases, 
they lead to unexpectedly large 
$B$(GT) and/or $B$(F) values. 
The same problem
has been often observed in other theoretical approaches
suggesting the necessity  of a 
quenching factor for $g_A$. 
%\cite{zuffi2003,pirinen2015,mardones2016,syoshida2018}. 
On the other hand, we do not introduce such 
a quenching in the present study. 
The results for  the $\beta$-decay $\log{ft}$ values serve as a 
sensitive test for the employed theoretical method, and 
may indicate certain improvements of the 
method, including the descriptions of the low-lying structures for 
both the even-even core and odd-mass parent and daughter nuclei. 
Possible ways to do so are the inclusions of 
new terms in the IBM and IBFM Hamiltonians or new 
degrees of freedom in the IBFM such as an intruder orbital 
coming from a next major shell. 
Another possibility to improve the description of the $\log{ft}$ values 
is to incorporate the effects of higher-order terms 
in the one-particle transfer operators in Eqs.~(\ref{eq:creation1}--\ref{eq:annihilation2}), 
as examined in Ref.~\cite{mardones2016}. 
This paper presents a first implementation 
of the EDF-based IBFM approach in the description 
of $\beta$ decay and, therefore, those extensions 
of the method are beyond the scope the present work 
and will be explored in the near future. 

%Even though only a 
%few coupling constants for the 
%boson-fermion interaction 
%terms in odd-A nuclei have to be determined empirically, 
%the essential ingredients of the 
%IBFM Hamiltonian (i.e., the even-even boson-core 
%Hamiltonian, the 
%single-particle energies and occupation probabilities 
%for the odd particle) have been determined microscopically  
%from EDF calculations, 
%and no additional phenomenological parameters have been
%used in the computation of the $B$(GT) and $B$(F) rates. 
In spite of the reduced number of empirical parameters in the model, it is possible to describe the 
detailed excitation spectra 
for even-even, odd-A, and odd-odd nuclear systems, 
and $\beta$-decay properties simultaneously and 
with reasonable computational time. 
Keeping all this in mind, we conclude that our
description of the low-lying states 
and  $\beta$ decay properties of odd-A nuclei within 
the EDF-based IBFM approach is fairly promising in the study of 
fundamental nuclear processes.

\begin{acknowledgments}
We thank N. Yoshida for helping us with the $\beta$ decay calculations  
The work of KN is financed within the Tenure Track Pilot Programme of 
the Croatian Science Foundation and the 
\'Ecole Polytechnique F\'ed\'erale de Lausanne, and 
the Project TTP-2018-07-3554 Exotic Nuclear Structure and Dynamics, 
with funds of the Croatian-Swiss Research Programme. 
The  work of LMR was supported by Spanish Ministry of Economy and Competitiveness (MINECO)
Grants No. PGC2018-094583-B-I00.
\end{acknowledgments}

\bibliography{refs}

\end{document}